\documentclass[
 preprint,
 amsmath,amssymb,
 aps,
]{revtex4-1}

\usepackage{graphicx}
\usepackage{dcolumn}
\usepackage{bm}

\begin{document}


\title{Universality of citation distributions and its explanation \\}

\author{Michael Golosovsky}
\email{michael.golosovsky@mail.huji.ac.il}
\affiliation{The Racah Institute of Physics, The Hebrew University of Jerusalem, 9190401 Jerusalem, Israel\\
}%
\date{\today}
\begin{abstract}
Universality or near-universality of citation distributions was found empirically  a decade ago but its theoretical justification has been lacking so far. Here, we systematically study citation distributions for different disciplines in order to characterize this putative  universality  and to understand it theoretically. Using our calibrated model of citation dynamics, we find microscopic explanation of the universality of citation distributions and explain deviations therefrom. We demonstrate that citation count of the paper is determined, on the one hand, by its fitness- the attribute  which, for  most  papers, is set at the moment of publication. The fitness distributions  for different disciplines are very similar and can be approximated by the log-normal distribution. On another hand,  citation dynamics of a paper is related to the mechanism by which  the knowledge about it spreads in the scientific community. This viral propagation is non-universal and discipline-specific. Thus, universality of citation distributions traces its origin to the fitness distribution, while deviations from universality are associated with the discipline-specific citation dynamics of papers.
\begin{description}
\item[PACS numbers]01.75.+m, 02.50.Ey, 89.75.Fb, 89.75.Hc
\end{description}
\end{abstract}
\pacs{{01.75.+m, 02.50.Ey, 89.75.Fb, 89.75.Hc}}
\keywords{Suggested keywords}
\maketitle
\section{Introduction}
Science is an evolving network of researchers, projects, and publications. Citations of scientific publications  are the most important links  that glue the whole network together.  Analysis of papers' citations  is important not only for bibliometrics  but for the whole field of complex networks \cite{Barabasi2015}, of which  citation networks were historically the first examples \cite{Price1976}.  Initially, this analysis focused on journal-based citation distributions. Although  these  distributions vary from discipline to discipline and from journal to journal, Seglen \cite{Seglen1992} established that, after proper scaling,  citation distributions  for different journals collapse onto a single curve. Radicchi, Fortunato, and Castellano  \cite{Radicchi2008} validated this observation and extended it to different disciplines, claiming that citation distributions are nearly universal. This claim was particularly telling to physicists, with their inclination to search for the universal laws of nature. It provided a stimulus  to look for universalities in other complex networks, and, indeed, several dynamic universalities  were found there as well \cite{Barzel2013,Gao2016,Candia2019}. While significant progress in  their understanding   has been achieved, the origin of the universality of citation distributions remained elusive. In the context of science of science \cite{Fortunato2018}, the striking  observation of Ref. \cite{Radicchi2008}  implies that different research topics develop along   similar paths, thus rendering possible such  generalizations  as the Kuhn's paradigm shift theory \cite{Kuhn1970}. The understanding of the universality of citation distributions may provide a solid base for  the Kuhn's theory which has been considered so far more like a philosophical idea rather than a scientific hypothesis.

Many empirical studies in the direction laid down by Radicchi et al. \cite{Radicchi2008}  were  performed by information scientists \cite{Bornmann2009,Waltman2011,Evans2012,Chatterjee2016}. In this area of research, requirements for  analysis are much more stringent than those accepted by  physicists, since the  primary motivation of the information scientists is to find a fair indicator allowing quantitative comparison of  the performance of  papers belonging to different scientific disciplines. In the language of information science, the main achievement of Ref. \cite{Radicchi2008}  is the demonstration that the variability of citation distributions for different fields is significantly reduced when one considers scaled citation distributions,   the scaling parameter being the mean of the distribution. The encompassing  studies  of Waltman et al. \cite{Waltman2011} showed deviations from the scaling suggested by Ref. \cite{Radicchi2008}, especially for the fields with low mean number of citations, and thus the limits to the claim of universality have been established. Subsequent studies \cite{Evans2012,Chatterjee2016} extended the scaling conjecture of Ref. \cite{Radicchi2008} to the sets of publications belonging to different journals, institutions,  and even to Mendeley readerships \cite{DAngelo2019}.   In general, these works supported the purported universality  but with some limitations, namely, the fields with  high number of uncited papers showed significant deviations from the universal distribution. To account for these deviations,  the two-parameter scaling \cite{Radicchi2011,Radicchi2012}  was suggested  as well.

Another evidence for universality or near-universality  came from empirical studies of  the functional shape of citation distributions. While early studies  \cite{Barabasi2015} tended to fit them with the power-law dependence,  later studies \cite{Stringer2008,Radicchi2008,Thelwall2016a} favored the log-normal fit,
\begin{equation}
\rho(K)=e^{-\frac{(\ln K-\mu)^{2}}{2\sigma^{2}}},
\label{lognormal}
\end{equation}
where $K$ is the number of citations of a paper, $\mu$ characterizes the mean number of citations for the set of papers, and $\sigma$ is the width of the distribution. The shape of the log-normal distribution in the log-log scale is uniquely determined by $\sigma$ and does not depend on $\mu$. Thus, Radicchi et al. \cite{Radicchi2008} claimed that  citation distributions for several natural science disciplines  follow the log-normal dependence with the same $\sigma$.  Extensive study of citation distributions for different journals by Thelwall \cite{Thelwall2016a} also indicated that they can be described by the log-normal distribution with nearly the same $\sigma=1-1.2$. This is in line with the earlier study of Stringer et al. \cite{Stringer2008} who reported  log-normal citation distribution with $\sigma\sim 1$  for hundreds of journals (obviously, the values of $\sigma$ listed in Ref. \cite{Stringer2008}  shall be multiplied by $\ln{10}=2.3025$). Ref. \cite{DAngelo2019}  reports $\sigma\sim 1$  for Mendeley readerships, Ref. \cite{Chatterjee2016} reports $\sigma=1.18$ for  many journal-based and institution-based publications, Ref. \cite{Evans2012}  reports $\sigma=1.14$. Thus,   the log-normal fit  of citation distributions  for different journals, fields, and institutions yields almost the same width $\sigma=1-1.2$, in other words, the shape of citation distributions  is nearly universal.

 After  universality or near-universality of citation distributions has been  established for many scientific disciplines (with some caveats), we are in a better position to assess its origins. It should be noted that citation distributions are not stationary, they result from the spread in paper's static and dynamic attributes. The former  are set at the time of the publication and include the journal, institution, reputation of the authors, the length of the paper, its genre, etc. Dynamic attributes include the initial citation history of the paper (impact factor), its niche in the evolving scientific community, how well did it catch attention there, etc. To understand citation distributions, we need to assess the sources of  variability in both the  static and dynamic attributes of papers. In this  study we focus  on the latter.

 While there were many insightful models of citation dynamics of papers, they didn't address  the universality of citation distributions, in such a way that it remained an empirical observation  lacking theoretical foundation. Our goal is to find an explanation of this universality based on  our recently developed microscopic model of citation dynamics  \cite{Golosovsky2017,Golosovsky2019}  which can be traced to the recursive search model of Ref. \cite{Simkin2007}.  Our model was carefully calibrated using one discipline- Physics and one publication year -1984. Here, we apply it to account for citation dynamics of  the Economics and Mathematics papers published in the same year. We measure the corresponding dynamic parameters and functions for these disciplines, and compare them to those for Physics. Some of these parameters  turned out to be universal while others are not. Basing on our model and on measured parameters and functions for  these three  disciplines, we explain the universality of citation distributions, as well as the deviations therefrom.

\section{Universality of citation distributions}
In what follows we illustrate what is meant by the universality of citation distributions. Consider a set of   papers published in the same year $t_{0}$ and denote by $k_{j}(t)$ the number of citations that a paper $j$ from this set garners in year $t_{0}+t$. We denote by $\Pi(K,t)=\int_{K}^{\infty}\rho(\kappa)d\kappa$  the cumulative probability distribution for this set at year $t$  where  $\rho(\kappa)$ is the corresponding probability density function. Although citation distributions for different years $t$ are markedly different, after dividing  each distribution   by the mean number of cumulative citations, $M(t)= \int_{o}^{\infty}K\rho(K,t)dK$,   the scaled distributions $\Pi\left(\frac{K(t)}{M(t)}\right)$  collapse onto a single curve. This scaling was first reported  by Radicchi et al. \cite{Radicchi2008} and received the name of  the universality of citation distributions.
 \begin{figure}[ht]
\includegraphics*[width=0.32\textwidth]{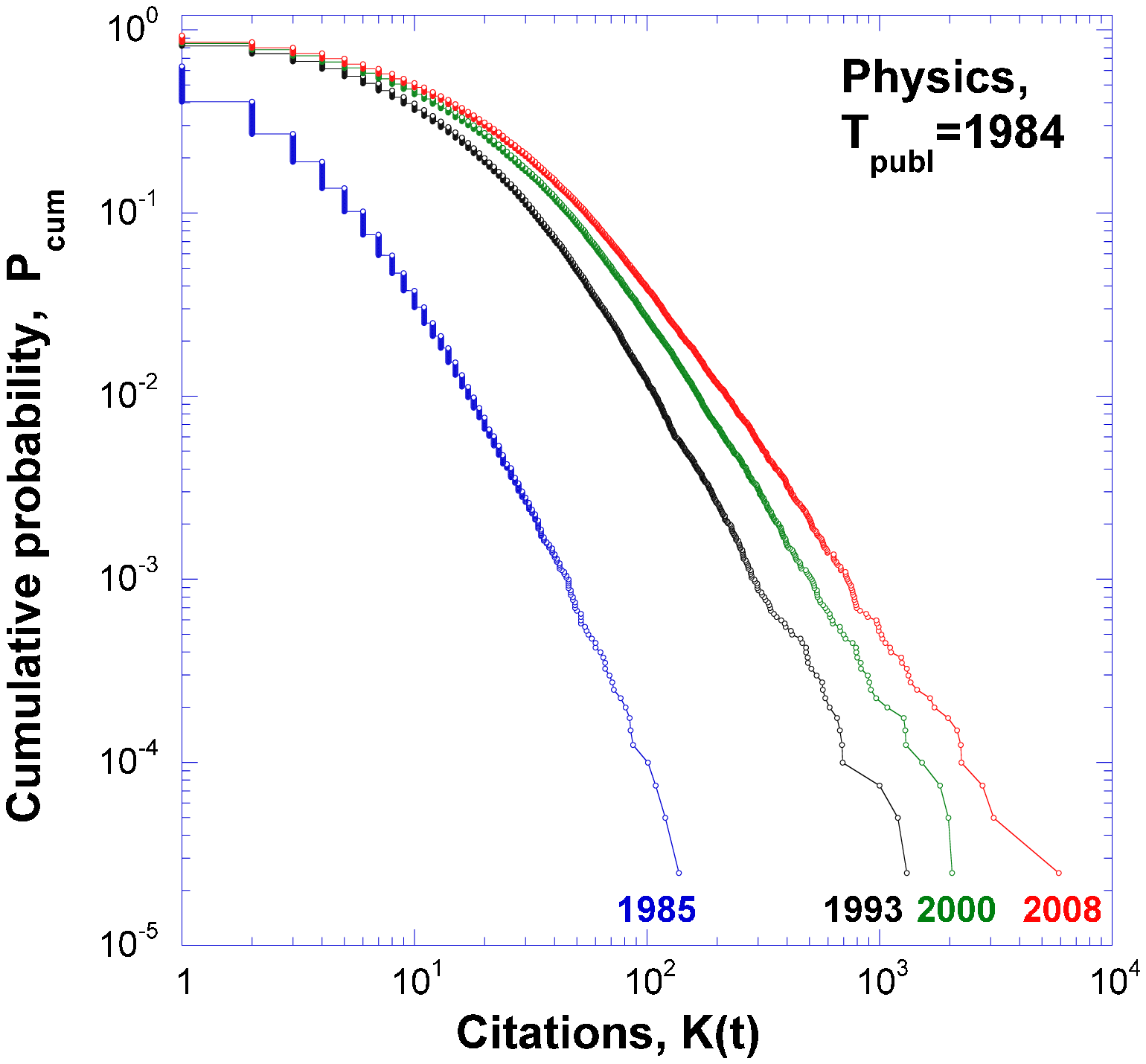}
\includegraphics*[width=0.32\textwidth]{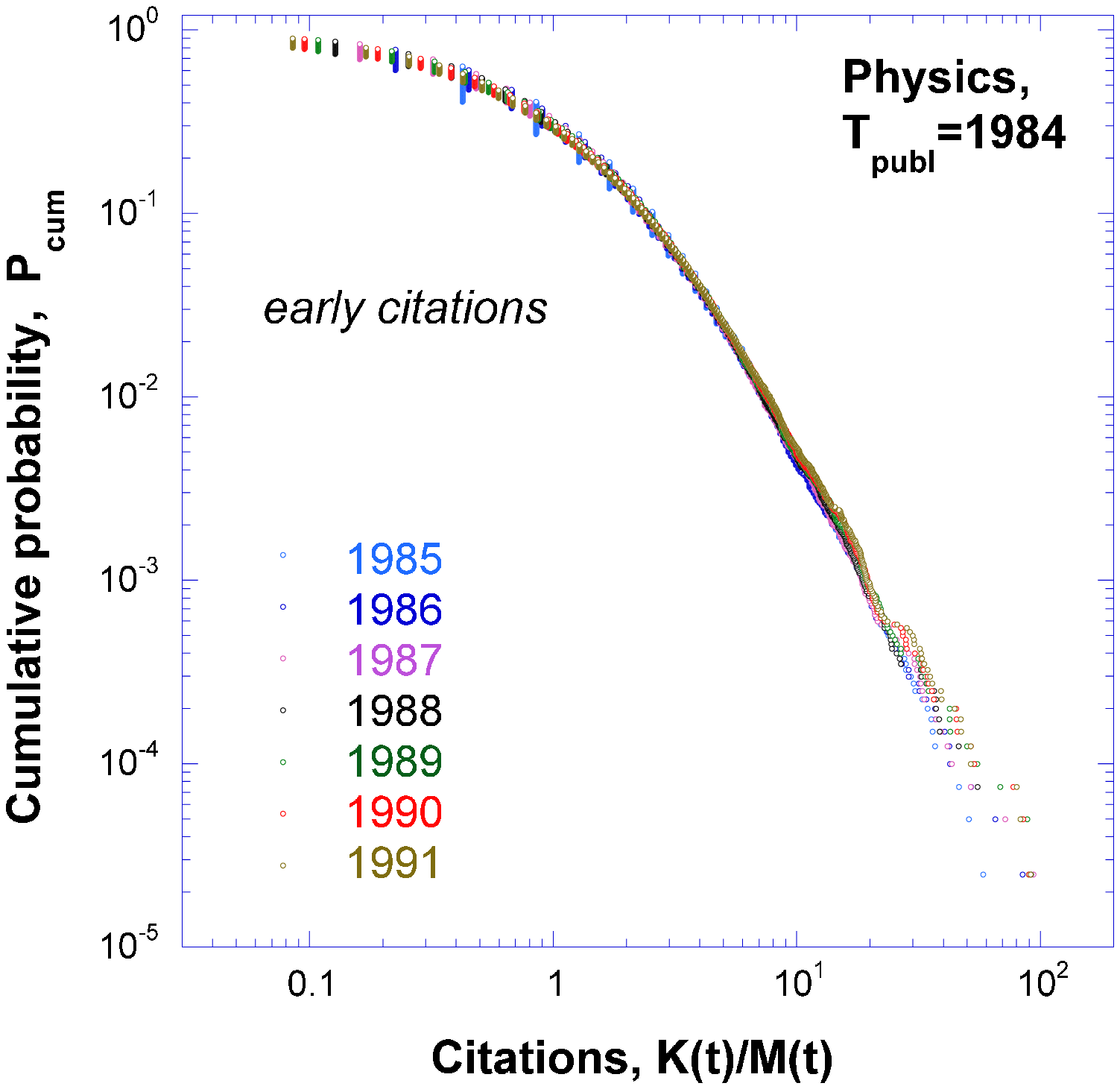}
\includegraphics*[width=0.32\textwidth]{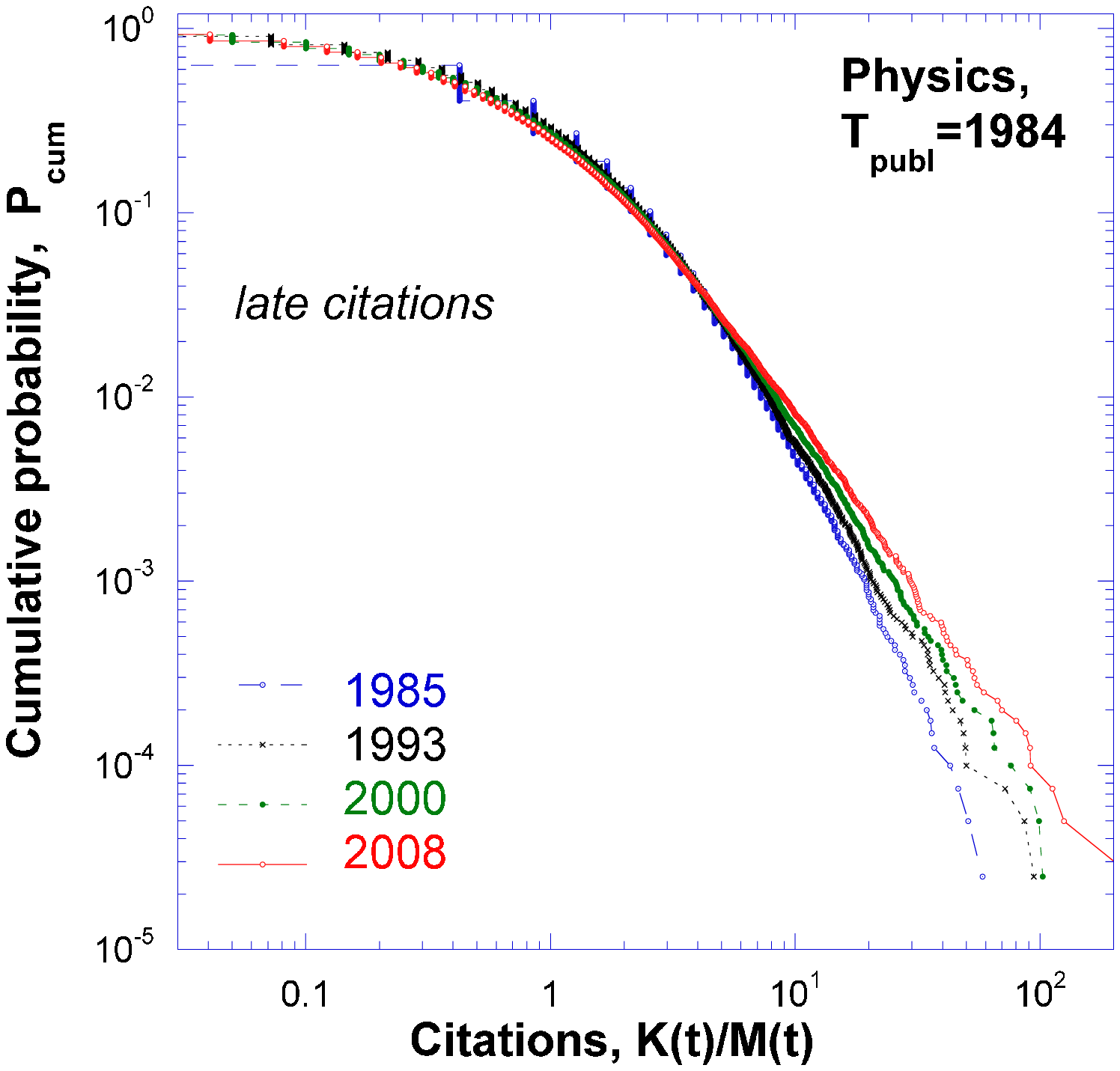}
\caption{ (a) Annual cumulative citation distributions, $\Pi(K)$,  for 40195 Physics papers published in 1984. (b) Scaled citation distributions, $\Pi\left(\frac{K(t)}{M(t)}\right)$, for early years. (c) Scaled citation distributions for late years. While early distributions  collapse onto a single  curve, late distributions show deviations, especially in their tails.
}
\label{fig:physics-years}
\end{figure}

Figure \ref{fig:physics-years}a  shows cumulative citation distributions  $\Pi(K,t)$ for all 40195 Physics papers published in 1984. Figures \ref{fig:physics-years}b and \ref{fig:physics-years}c show early ($t=1-6$ years after publication), and late ($t=$7-25 years after publication) scaled citation distributions, correspondingly.  While early distributions collapse onto a single curve, late distributions do not collapse well and show significant deviations  in their tails.  Thus, one-parameter scaling suggested in Ref. \cite{Radicchi2008} for the papers in one discipline published in the same  year, is valid only for early citation distributions. As time passes, the one-parameter scaling becomes unsatisfactory.

\begin{figure}[ht]
\includegraphics*[width=0.35\textwidth]{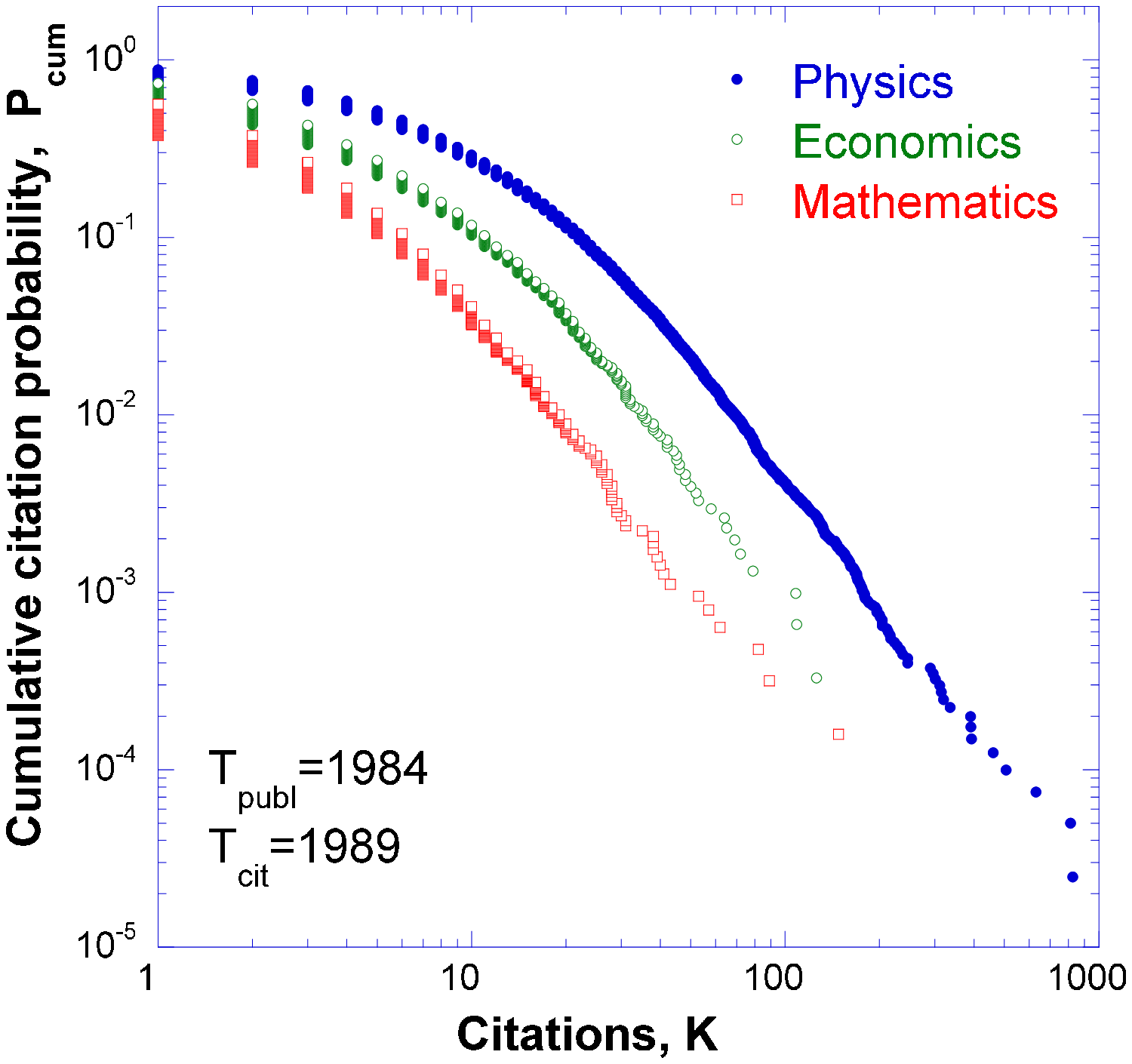}
\includegraphics*[width=0.35\textwidth]{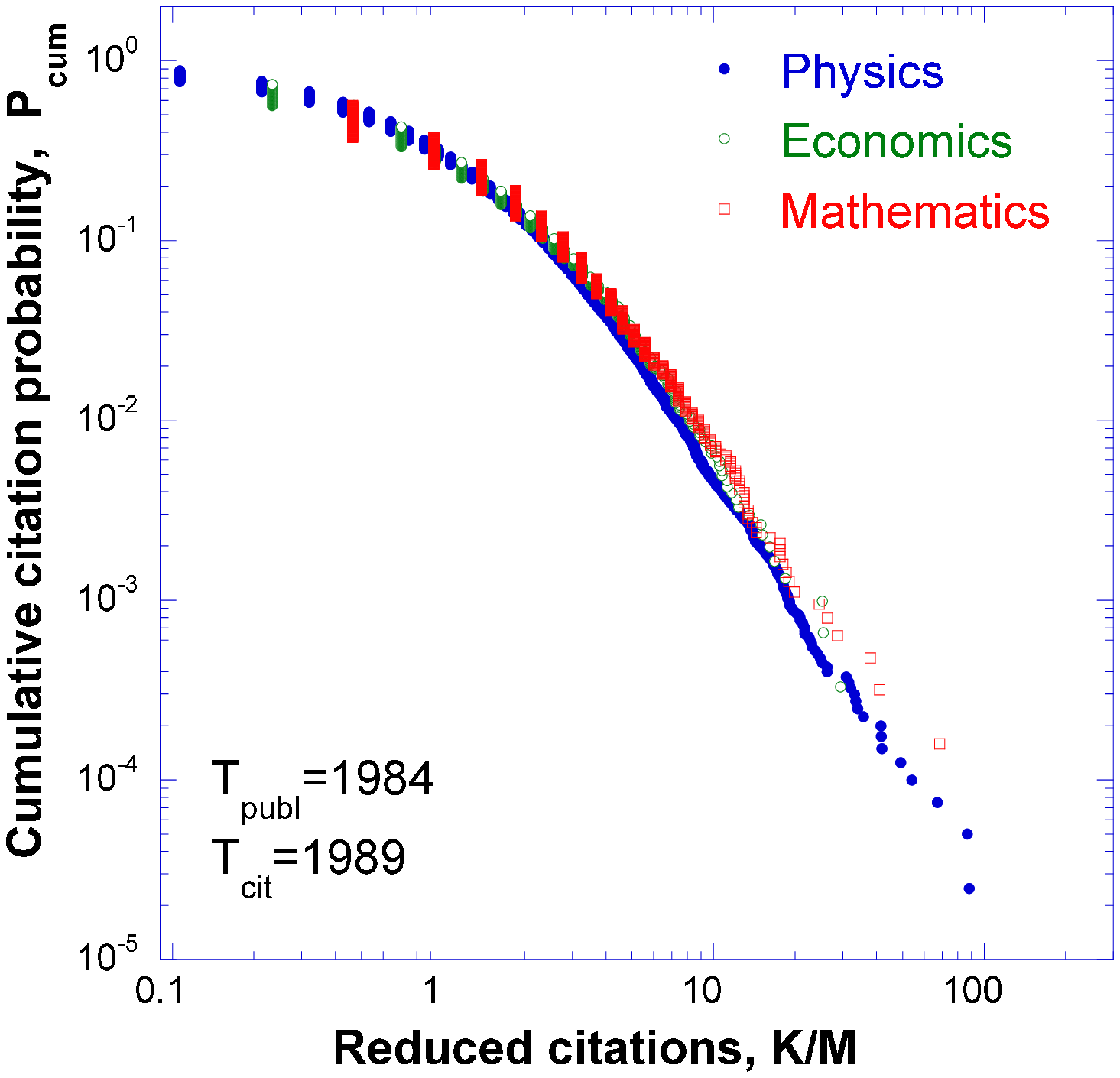}
\caption{(a) Annual cumulative citation distributions  $\Pi(K)$  for 40195 Physics papers, 6313 pure Mathematics papers, and 3043 Economics papers, all published  in 1984 and measured in 1990.   (b)  Scaled  distributions  $\Pi\left(\frac{K}{M}\right)$, where $M$ is the mean of the corresponding distribution. The  scaled distributions for all three disciplines collapse onto one  curve.
}
\label{fig:disciplines}
\end{figure}
\begin{figure}[ht]
\includegraphics*[width=0.35\textwidth]{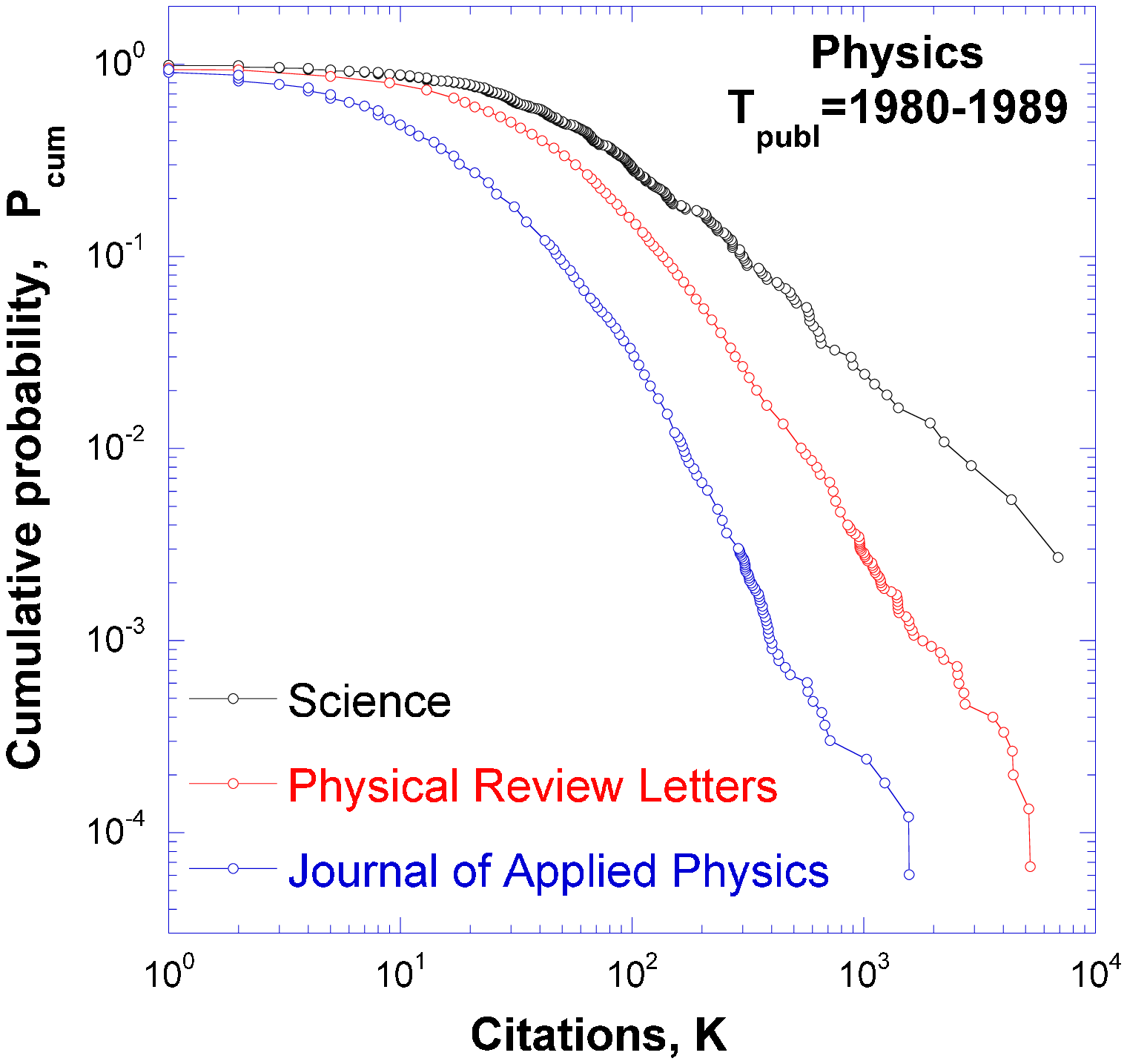}
\includegraphics*[width=0.35\textwidth]{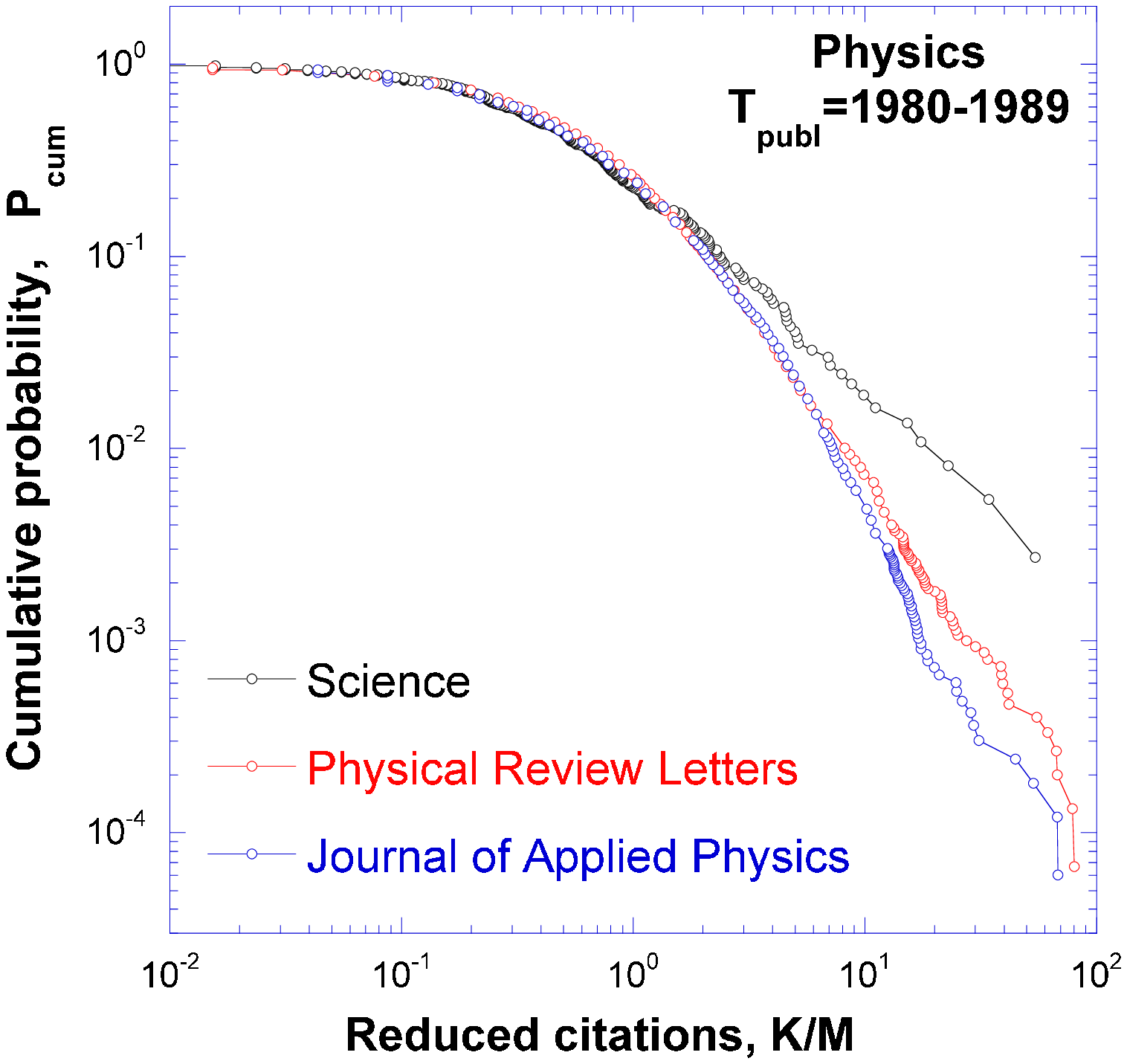}
\caption{Annual cumulative citation distributions for   Physics papers published in three different journals: Journal of Applied Physics, Physical Review Letters, and Science (only  Physics papers). Publication year is 1984, citations are counted 25 years after publication.  (a) Cumulative citation distributions,  $\Pi(K)$. (b)  Scaled  cumulative citations $\Pi\left(\frac{K}{M}\right)$, where $M$ is the mean of the corresponding distribution. Citation distributions for these journals do not scale.
}
\label{fig:physics-journals}
\end{figure}
Figure \ref{fig:disciplines} shows another example of scaling. Here, we compare citation distributions for the papers  belonging to three different disciplines and published in the same year.  When each distribution is divided by its mean, they collapse onto a single curve. Again,  this scaling works well for early citation distributions and breaks for late distributions (not shown here).

Figure \ref{fig:physics-journals} compares citation distributions for Physics papers published in three different journals in the same year. Obviously, these distributions are very different, and after dividing them by  the mean number of citations, they do not scale at all. Here, we intentionally focused on the  journals with very different mean number of citations. If we were considering the journals with more or less the same mean, these distributions would scale \cite{Stringer2008,Thelwall2016b}.

Thus, in some cases scaled citation distributions $\Pi\left(\frac{K}{M}\right)$ exhibit universality while in other cases they do not. In what follows we explain these observations using our model of citation dynamics \cite{Golosovsky2017,Golosovsky2019}.

\section{The model of citation dynamics based on recursive search}
We present here a short summary of this model. Consider some paper $j$ published in year $t_{j}$. The  author of a new paper published  $t$ years later,   may cite the paper $j$ after finding it in databases, in  scientific journals,  or following recommendations of colleagues or news portal. This process results in  direct citation. An author of another new paper can find the paper $j$ in the reference list of the paper already included in his reference list and cite it. This process results in  indirect citation.

We assume that citation dynamics of papers follows an inhomogeneous Poisson process, namely, the papers'  citation rate  in year $t$  has a probability distribution  $\frac{\lambda_{j}^{k_{j}}}{k_{j}!}e^{-\lambda_{j}}$ where $\lambda_{j}(t)$ is  the latent citation rate. The model  stipulates that $\lambda_{j}(t)$  is given by the following expression,
\begin{equation}
\lambda_{j}(t)=\eta_{j}R_{0}\tilde{A}(t)+\int_{0}^{t}m(t-\tau)P(K_{j},t-\tau)k_{j}(\tau)d\tau,
\label{lambda}
\end{equation}
 where the first  and the second addends correspond, respectively, to direct and indirect citations. Here,   $\tilde{A}(t)$ is the aging function for citations,  $t$ is the number of years after publication of paper $j$,   $R_{0}$ is the average length of the reference list of  the  papers  published  in the same year and belonging to the same discipline, and $\eta_{j}$ is the papers' fitness. It characterizes the appeal which the paper $j$ makes to the readers. Our definition of fitness is different from that of Ref. \cite{Bianconi2001}  and goes after Caldarelli et al. \cite{Caldarelli2002}, in particular, we assume that $\eta_{j}$  does not change during the papers' lifetime.


Each direct citation of the paper $j$ triggers  cascades of indirect citations which are captured by the second  addend in Eq. \ref{lambda}.  Here,  $k_{j}(\tau)$ is the number of  citations garnered  by the paper $j$   in  year $\tau$ (the number of the first-generation citing papers);  $m(t-\tau)$ is the average number of the (second-generation) citing papers  garnered by a first-generation citing paper   in year $t$ (we disregard here the difference between the number  of the second-generation  citations and citing papers which we discussed in detail  in  \cite{Golosovsky2017,Golosovsky2019});  $K_{j}$ is the cumulative number of citations of the paper $j$, and $P(K_{j},t-\tau)$ is the probability of a second-generation citing paper  to cite paper $j$. Our measurements  for Physics papers yielded that this probability quickly decays with time,
\begin{equation}
P(K,t-\tau)=P_{0}e^{-\gamma(t-\tau)},
\label{P}
\end{equation}
where $\gamma$ is the obsolescence rate and $P_{0}$ is the probability magnitude  which slowly increases with the number of accumulated citations $K$,
\begin{equation}
P_{0}(K)=\frac{T_{0}}{R_{0}}(1+b\ln K).
\label{T}
\end{equation}
$T_{0},b$  and $\gamma$ are empirical parameters,  the same for all papers in one discipline,  and the index $j$ was dropped for clarity. Notably, the $P_{0}(K)$ dependence introduces nonlinearity into  Eq. \ref{lambda}, its magnitude being characterized by the nonlinear coefficient $b$.

Equations \ref{lambda}-\ref{m} indicate that  citation dynamics of a paper $j$   is determined by its fitness $\eta_{j}$, by the aging function $\tilde{A}(t)$, and by the parameters $\eta_{0}R_{0},\frac{T_{0}}{R_{0}},b,\gamma$  which are  common for all papers in one discipline. The composite parameters $\eta_{0}R_{0}$ and $\frac{T_{0}}{R_{0}}$  include the factor $R_{0}$, the average  reference list length. We introduced it already at this stage since it defines the scale  for the average citation rate of papers belonging to one discipline  and, as we will show soon, it is the dominant factor which determines the difference between citation distributions for different disciplines.
$R_{0}$ is closely related to the mean citation rate, $m(t)=\overline{k_{j}}(t)$. The latter is a solution of Eq. \ref{lambda} averaged over all papers in the same discipline published in the same year, namely,
\begin{equation}
m(t)=\eta_{0}R_{0}\tilde{A}(t)+\int_{0}^{t}m(t-\tau)\overline{P_{0}}e^{-\gamma(t-\tau)} m(\tau)d\tau,
\label{m}
\end{equation}
where $\eta_{0}$  is the average fitness and $\overline{P_{0}}$ is the average probability of indirect citation.

Since one's paper citation is another paper's reference, there is a certain duality between the references and citations, whereas the analog of $m(t)$ is $R(t)$, the age distribution of references, namely,  the average number of references  of age $t$ in the reference list of  papers belonging to one discipline and published in one year $t_{0}$.  In particular,
\begin{equation}
R(t)=m(t)e^{-(\alpha+\beta)t},
\label{duality}
\end{equation}
where  the  number of publications $N$ and the average reference list length $R_{0}$ are assumed to  grow  exponentially \cite{Golosovsky2017,Golosovsky2019}, $N\propto e^{\alpha t_{0}}$, $R_{0}\propto e^{\beta t_{0}}$.   Equation \ref{duality} yields that  $R_{0}=\int_{0}^{t}R(\tau)d\tau= \int_{0}^{t}m(\tau)e^{-(\alpha+\beta) \tau}d\tau$.

Further consideration of $R(t)$ allows us to find relation between different parameters of citation dynamics. Indeed,  Eqs.  \ref{m}, \ref{duality} yield
\begin{equation}
R(t)=\eta_{0}R_{0}A(t)+
\int_{0}^{t}R(\tau)\frac{\overline{T}}{R_{0}}e^{-\gamma(t-\tau)}R(\tau)d\tau,
\label{R}
\end{equation}
where
\begin{equation}
A(t)=\tilde{A}(t)e^{-(\alpha+\beta)t}
\label{A}
\end{equation}
 is the aging function for references  which has been defined in such a way as to obey  normalization condition, $\int_{0}^{\infty}A(\tau)d\tau=1$ \cite{Golosovsky2017}.   We also introduce  the reduced age distribution of references, $r(t)=\frac{R(t)}{R_{0}}$, which is better known as synchronous or retrospective  citation distribution \cite{Nakamoto1988,Glanzel2004}.  Equation \ref{R} yields
\begin{equation}
r(t)=\eta_{0}A(t)+\int_{0}^{t}r(\tau)\overline{T}e^{-\gamma(t-\tau)}r(t-\tau)d\tau.
\label{r}
\end{equation}
Equation \ref{r}  is more general than the parent Eq. \ref{R} and it does not contain $R_{0}$. The reason to introduce the composite parameters $\eta_{0}R_{0}$ and $\frac{T_{0}}{R_{0}}$  into Eq. \ref{lambda}   was  to get rid of $R_{0}$ in Eq. \ref{r}.  Integration of  Eq. \ref{r} over time yields
\begin{equation}
1=\eta_{0}+\overline{T}\int_{0}^{\infty}r(\tau)e^{-\gamma\tau}d\tau.
\label{rr}
\end{equation}
This equation connects together several parameters related to indirect citations. Indeed, Roth et al.  \cite{Roth2012}  showed that, after proper time rescaling $\tau\rightarrow\tilde{\tau}$, the $r(\tilde{\tau})$ dependences for different disciplines nearly collapse onto one curve. Being motivated by this observation, we  introduce rescaled time $\tilde{\tau}=\gamma\tau$ and recast Eq. \ref{rr} as
\begin{equation}
\int_{0}^{\infty}r(\tilde{\tau})e^{-\tilde{\tau}}d\tilde{\tau}=(1-\eta_{0})\frac{\gamma}{\overline{T}}.
\label{rrr}
\end{equation}
 Following observation of Ref. \cite{Roth2012}, we speculate that  $r(\tilde{\tau})$ dependences for different disciplines are very similar.   Then Eq. \ref{rrr}  implies that the parameters $\eta_{0},\gamma$, and $\overline{T}$ are not independent, in particular, the factor $(1-\eta_{0})\frac{\gamma}{\overline{T}}$ should be discipline-independent.

After such thorough discussion of the parameters defining citation dynamics of individual papers, we switch to citation distribution for a collection of papers, all published in the same year. Our model yields that the cumulative citation distribution  for a collection of papers is $\Pi_{i}(K,t)=\int_{0}^{\infty}K(\eta,t)\rho_{i}(\eta)d\eta$, where  $K(\eta,t)$ is the mean number of citations garnered by the  papers with the same fitness $\eta$,  $t$ is the number of years after publication, and $\rho_{i}(\eta)$, is the fitness distribution for this collection.  In our previous study we measured  the parameters  that determine  citation dynamics of Physics papers  and the corresponding citation distributions \cite{Golosovsky2017}.   Here, we determine the corresponding parameters and functions  for  Economics and Mathematics papers as well, compare them to those for Physics, and  decide which of them are universal and which are not.
\section{Model calibration for Mathematics and Economics papers}
\subsection{Analysis of citation distributions}
\begin{figure}[ht]
\begin{center}
\includegraphics*[width=0.35\textwidth]{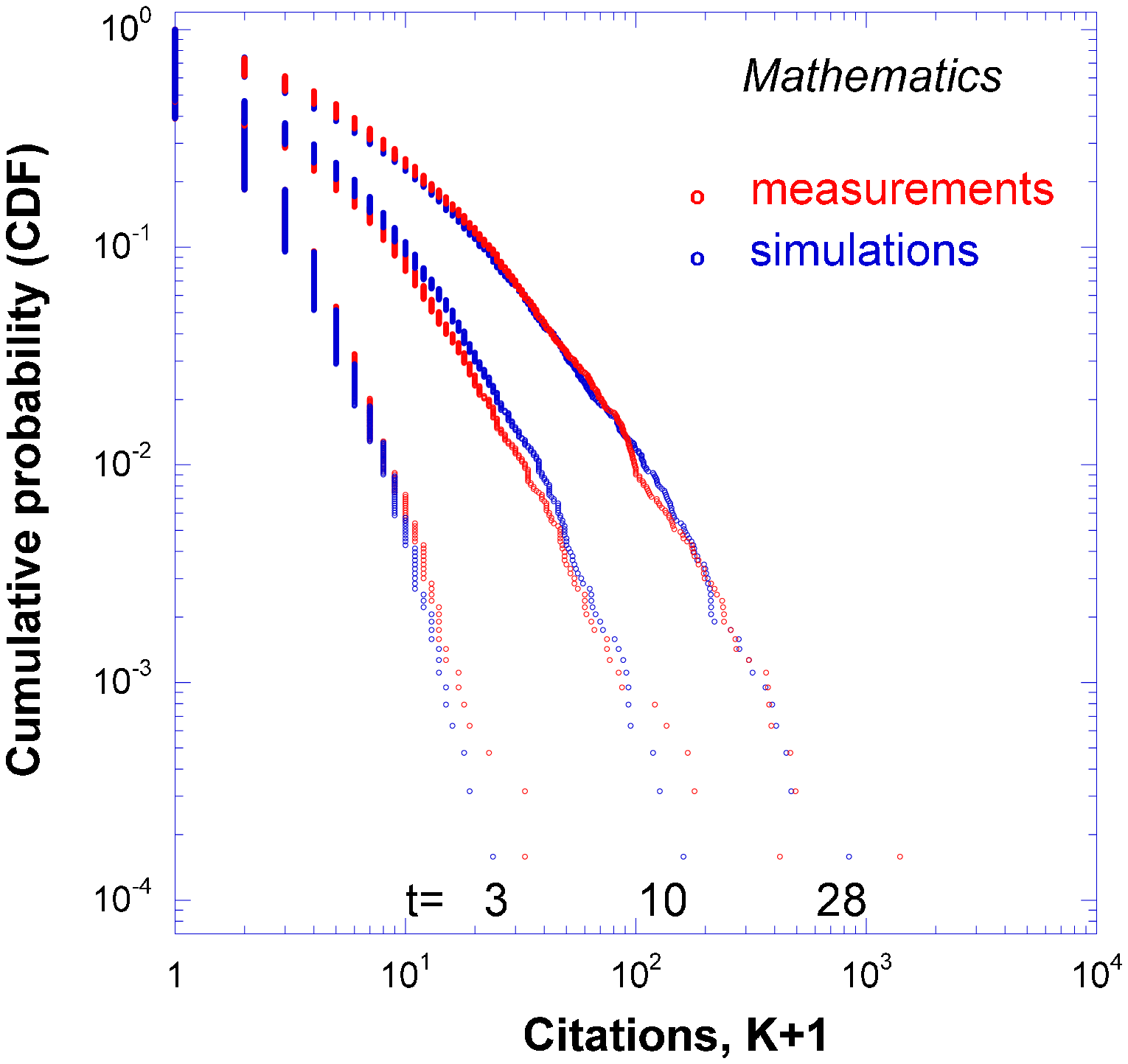}
\includegraphics*[width=0.35\textwidth]{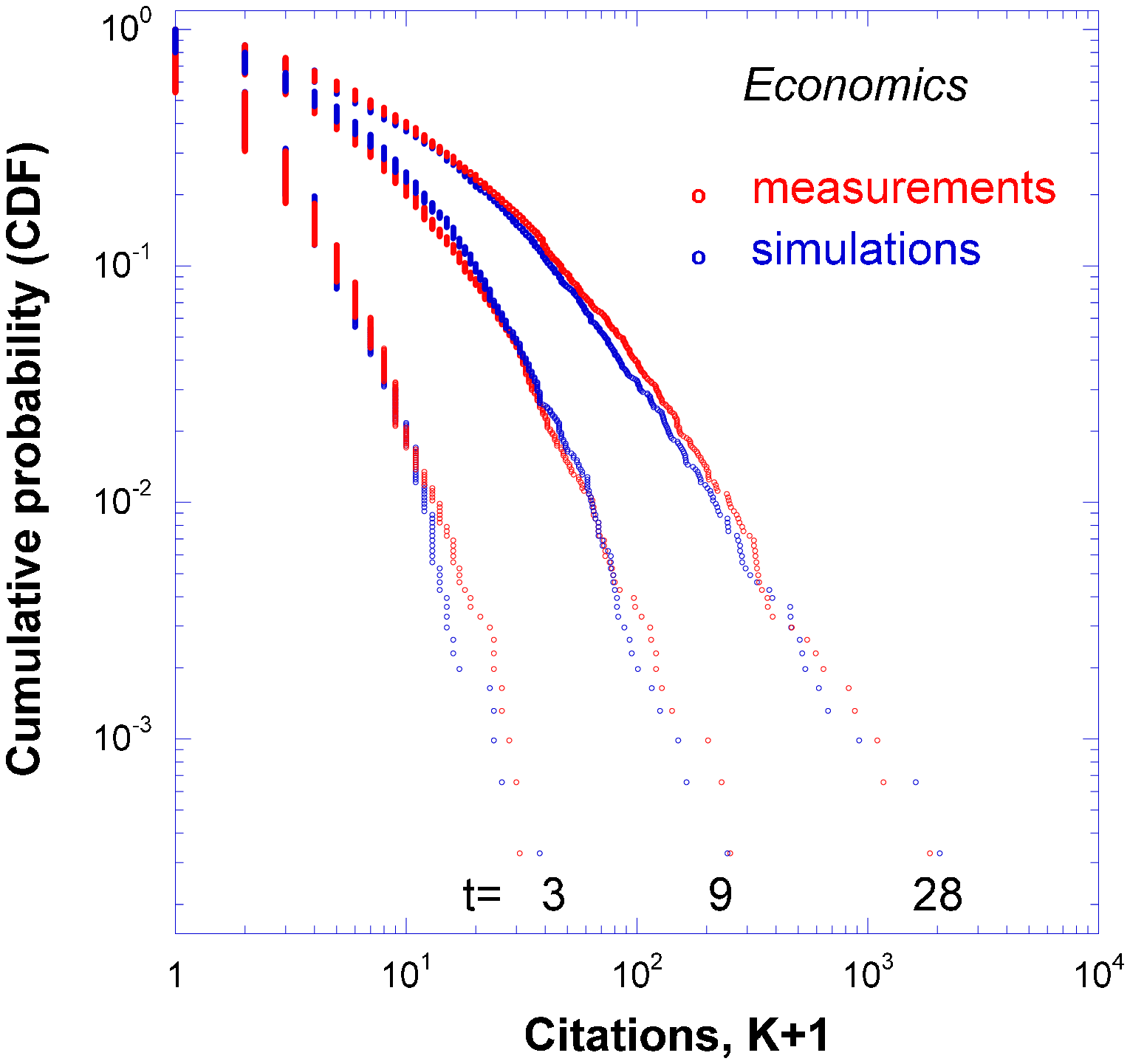}
\caption{Annual cumulative citation distributions for  6313 Mathematics papers published in 1984 (left panel) and 3043  Economics papers published in 1984 (right panel).  Red circles show  measured data, blue circles show results of stochastic simulation based on the Poisson process with the rate given by Eq. \ref{lambda}.    $t$ is the time after publication, the publication year corresponding to $t=1$.
}
\label{fig:distributions}
\end{center}
\end{figure}

 Using Clarivate WoS, we pinpointed all pure Mathematics and all Economics papers published in one year- 1984, and measured their citation dynamics during subsequent 28 years. We included in our analysis research papers, letters, and notes (and uncited papers as well) and excluded reviews and editorial material.  Figure \ref{fig:distributions} shows measured  citation distributions.
To compare them to our model, we  ran  stochastic numerical simulation based on Eq. \ref{lambda}. To find the combination of the fitting parameters providing the best fit to  measured citation distributions,  we used a stepwise  fitting procedure.  First, we analyzed the mean number of citations $M(t)$ and determined the average reference list length $R_{0}$. Secondly, we crudely estimated  the fitness distribution $\rho(\eta)$ from the comparison of the early citation distributions to numerical simulation, basing on the fact that they  mimic fitness distribution because they consist mostly of direct citations.  Third, we substituted the fitness distribution  into our model and fitted  both early and late citation distributions using  $\gamma, T_{0},b$  as fitting parameters and  $\tilde{A}(t)$ as a fitting function. Then the  latter was fine-tuned to closely match the $M(t)$ dependence while $\gamma,T_{0}$, and $b$ were fine-tuned to  fit the tails of the citation distributions.

If citation distributions were the only way of comparison between the model and the data, this procedure would be still inconclusive  since several combinations of the fitting parameters can produce similar citation  distributions. To lift this ambiguity, we performed  cross checks and, in addition to citation distributions,  considered  those aspects of citation dynamics that are insensitive to the fitness distribution and to the aging function. In particular, we analyzed the Pearson autocorrelation coefficient which characterizes fluctuations of the papers' citation trajectory.  It is defined as follows. Consider a subset of papers that  garnered the same number of  citations $K(t)$ during $t$ years after publication. For  this subset, we determine  the number of citations garnered by each paper  during two subsequent years,   $k_{j}(t)$ and $k_{j}(t-1)$. The Pearson autocorrelation coefficient is
\begin{equation}
c_{t,t-1}(K)=\frac{\overline{\left(k_{j}(t)-\overline{k_{j}}(t)\right)\left( k_{j}(t-1)-\overline{k_{j}}(t-1)\right)}}{\sigma_{t}\sigma_{t-1}},
\label{c}
\end{equation}
where  $\overline{k_{j}}(t)$ and $\overline{k_{j}}(t-1)$  are, correspondingly, the mean of the $k_{j}(t)$ and $k_{j}(t-1)$ distributions, and $\sigma(t)$ and $\sigma(t-1)$  are their standard deviations.

 Close inspection of Eqs. \ref{lambda}, \ref{c} indicates that $c_{t,t-1}$ is insensitive to $\tilde{A}(t)$ and $\rho(\eta)$. Moreover, $c_{t,t-1}=0$ for direct citations   since their statistics is Poissonian, in such a way that the direct citation rates of a paper in subsequent years are uncorrelated.  The autocorrelation coefficient is mostly sensitive to indirect citations which are determined by the previous citation history of a paper, in particular,  $c_{t,t-1}=1$ indicates that citation rate of a paper is fully determined by the number of citations garnered last year. Thus, $c_{t,t-1}$ is sensitive only to parameters  $\gamma,T_{0},b$ which characterize indirect citations. By fitting the measured and numerically simulated $c_{t,t-1}(K)$ dependences we determine these parameters.

In the same vein, we analyzed  the paper's citation lifetime $\tau_{0}=\Gamma^{-1}$ which  is defined from the exponential approximation of the papers' citation trajectory, $K(t)=K_{\infty}(1-e^{-\Gamma t})$ where   $\tau_{0}=1/\Gamma$. Our measurements \cite{Golosovsky2017a} yielded that $\Gamma$ decreases with  $K$
\begin{equation}
\Gamma=\Gamma_{0}-G\ln K.
\label{G}
\end{equation}
The decreasing $\Gamma(K)$ dependence results in  a spectacular phenomenon-  citation lifetime of a paper increases with the number of citations  in such a way that highly-cited papers become runaways \cite{Golosovsky2012a}. This is a direct consequence of the nonlinear Eq. \ref{lambda}, where the nonlinearity is  introduced through Eq. \ref{T}. The empirical coefficient $G$ in  Eq. \ref{G} is closely related to the parameter $b$ in Eq. \ref{T} since the latter is the only source of  nonlinearity in Eq. \ref{lambda}. In our fitting procedure, we verified that the  the numerically-simulated and measured $G(b)$ dependences match one another.

Thus, for each discipline we found such combination of the fitting parameters that allows for our numerical simulation to closely fit not only the measured citation distributions  $\Pi(K,t)$ but the time-dependent  mean citation rate $m(t)$, the  Pearson autocorrelation coefficient $c_{t,t-1}(K)$, and the inverse citation lifetime $\Gamma (K)$ as well.

 \subsection{Parameters of citation dynamics for different disciplines}
 \subsubsection{Mean number of citations, $M(t)$}
\begin{figure}[ht]
\includegraphics*[width=0.35\textwidth]{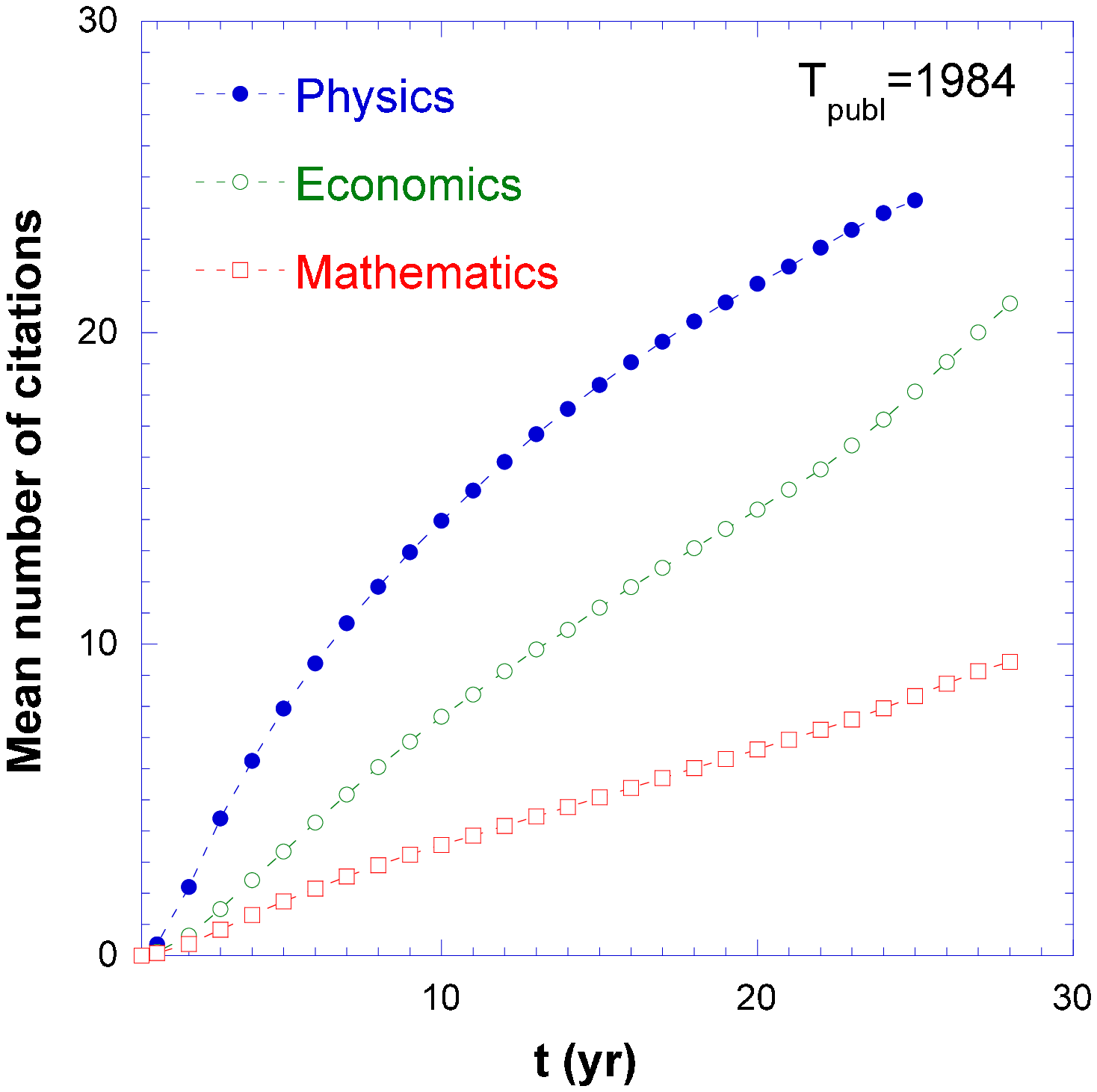}
\includegraphics*[width=0.35\textwidth]{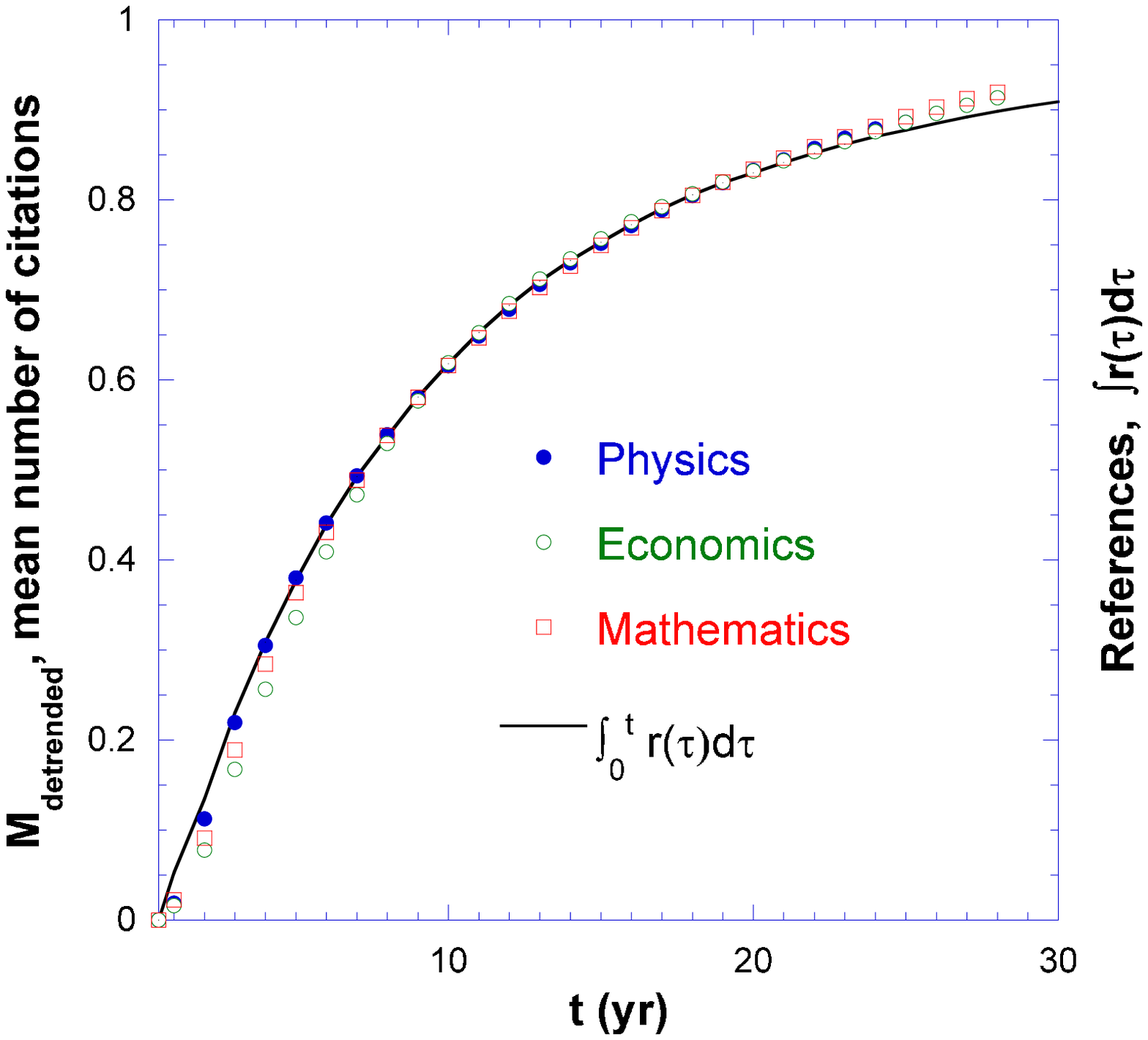}
\caption{The mean number of citations for  40195 Physics,  6313 Mathematics, and 3043 Economics papers, all published in 1984. (a) Raw data.  For   Physics papers, $M(t)$ grows with deceleration, while for Mathematics and Economics papers, $M(t)$ exhibits a permanent growth with no signs of saturation. (b) Detrended and scaled data, $\frac{M_{detrended}(t)}{R_{0}}=\int_{0}^{t}\frac{m(\tau)}{R_{0}}e^{-(\alpha+\beta)\tau}d\tau$  where $R_{0}$ is the average reference list length for each discipline.  The data for three disciplines  nearly collapse onto one curve. Continuous line shows   $\int_{0}^{t}r(\tau)d\tau$ dependence measured directly from the analysis of the age composition of the  reference lists of Physics papers \cite{Golosovsky2017}. It is almost identical to   $\frac{M_{detrended}(t)}{R_{0}}$ dependence, as expected from Eq. \ref{r}. 
}
\label{fig:mean}
\end{figure}

 Figure \ref{fig:mean}  shows the time dependence of the mean number of cumulative citations  for three disciplines. These dependences are different: in the long time limit, they either show signs of  saturation or diverge. We attribute this divergence  to the exponential growth of the number of publications and of the reference list length. To put different $M(t)$ dependences on common ground, we recur to Eq. \ref{duality} and consider $M_{detrended}(t) =\int_{0}^{t}m(\tau)e^{-(\alpha+\beta) \tau}d\tau$ where  $\alpha$ and $\beta$ are the exponents, characterizing the exponential growth of the number of papers, and of the reference list length, correspondingly.  According to our model, $M_{detrended}(t)$ is nothing else but $R(t)$, the age composition of the reference list which has been discussed in relation to Eq. \ref{R}.  While $M(t)$ can diverge in the long time limit, $M_{detrended}(t)$ converges to  the average reference list length $R_{0}$. Notably, $\frac{M_{detrended}(t)}{R_{0}}=r(t)$, the reduced age distribution of references (see Eq. \ref{r}).

For each discipline,  we found  by trial and error the sum of the growth exponents $\alpha+\beta$ and the reference list length $R_{0}$   from the condition that $\frac{M_{detrended}(t)}{R_{0}}$ converges to unity in the long time limit. For  Physics, Economics, and Mathematics papers published in 1984, this yields,  $\alpha+\beta=0.04, 0.085, 0.092$ yr$^{-1}$; and $R_{0}=18, 8, 3.6$, correspondingly.  The value of $R_{0}$ for Physics is close to that found in our direct  measurements \cite{Golosovsky2017}, while $R_{0}$  values for Economics and Mathematics are underestimated. It should be noted, however, that  $R_{0}$  defined through Eq. \ref{duality} includes only original research papers and excludes books, conference proceedings, and interdisciplinary references.  These constitute a very small part of Physics references while they are abundant among  Mathematics and Economics references, hence the effective $R_{0}$ for these disciplines is smaller than the actual reference list length.

\subsubsection{Aging function, $\tilde{A}(t)$}
\begin{figure}[ht]
\includegraphics*[width=0.35\textwidth]{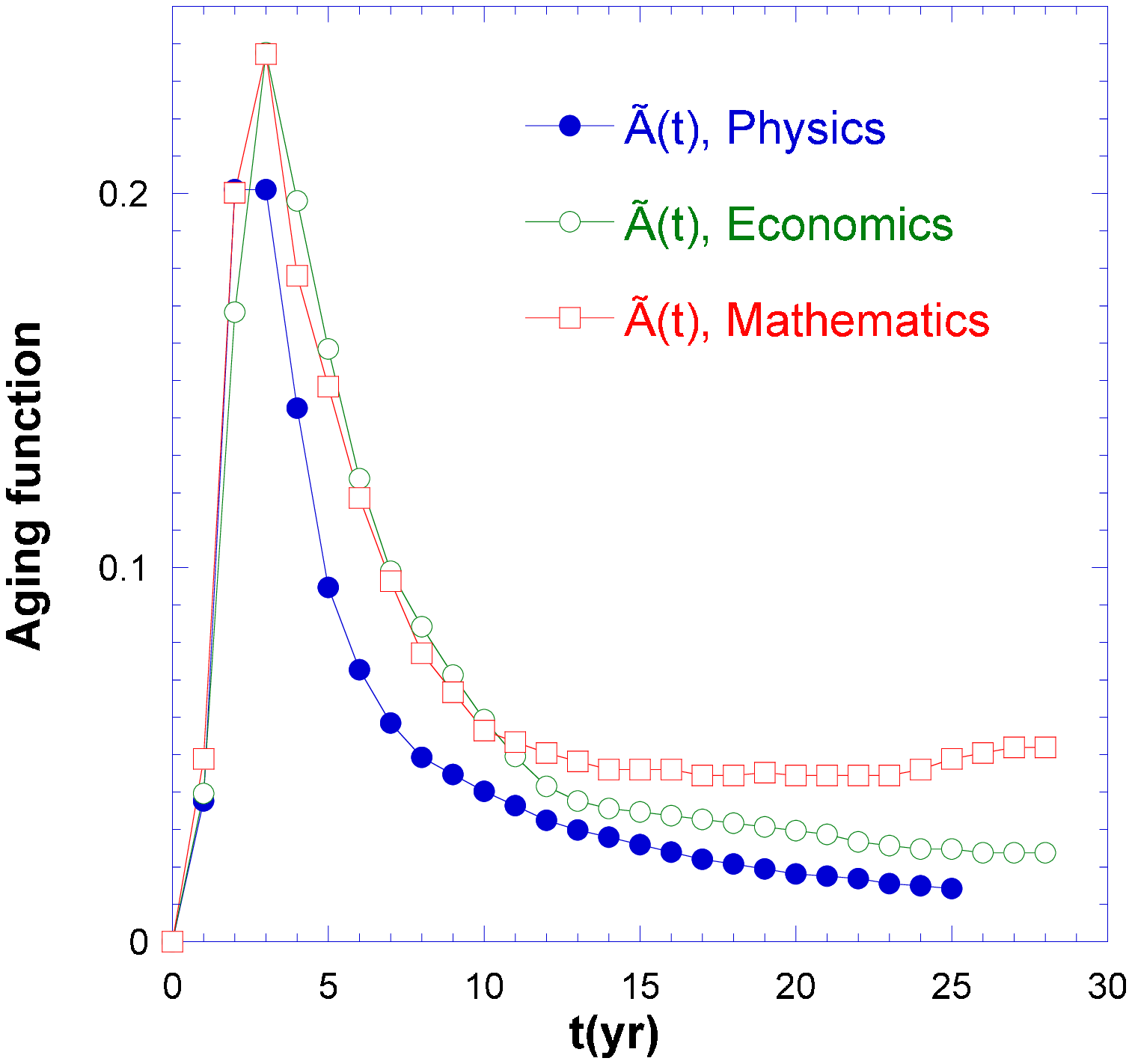}
\includegraphics*[width=0.35\textwidth]{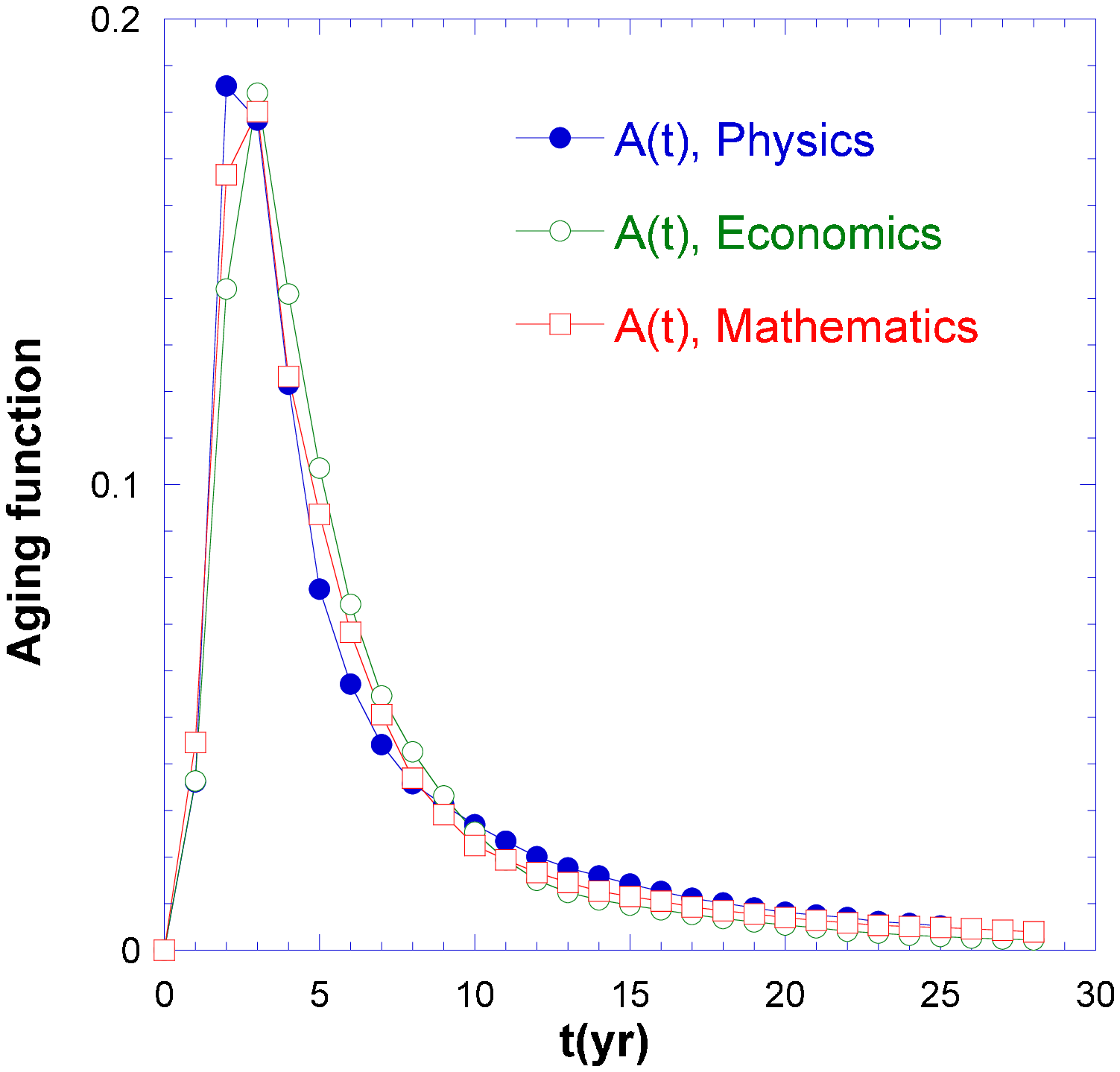}
\caption{ (a) $\tilde{A}(t)$, aging functions for citations as  found from the analysis of citation distributions of the  Physics, Mathematics, and Economics papers published in 1984.  (b) Detrended aging function $A(t)$,  as found from  Eq. \ref{A}.  While $\tilde{A}(t)$ dependences are different, $A(t)$ dependences  for these three disciplines collapse onto a single curve.
}
\label{fig:aging-function}
\end{figure}

Figure \ref{fig:aging-function}a shows that  the aging functions for citations  differ from
discipline to discipline. We found that this difference stems from the discipline-specific growth exponents, $\alpha$ and $\beta$.  To demonstrate this, we considered the detrended aging function $A(t)=\tilde{A}(t)e^{-(\alpha+\beta) t}$ which is nothing else but the aging function for references (Eq. \ref{A}). We  substituted into Eq. \ref{A} the sum of the growth exponents $\alpha+\beta$ determined from the Fig. \ref{fig:mean} and found $A(t)$ for each discipline. Figure \ref{fig:aging-function}b shows that all three $A(t)$ dependences  collapse onto one curve. Although  in the framework of our model,  $A(t)$  dependence  could be discipline-specific, Figure \ref{fig:aging-function}b proves that it is universal, at least for three widely different disciplines of our study.

 \subsubsection{Fitness distribution $\rho(\eta)$}
\begin{figure}[ht]
\includegraphics*[width=0.35\textwidth]{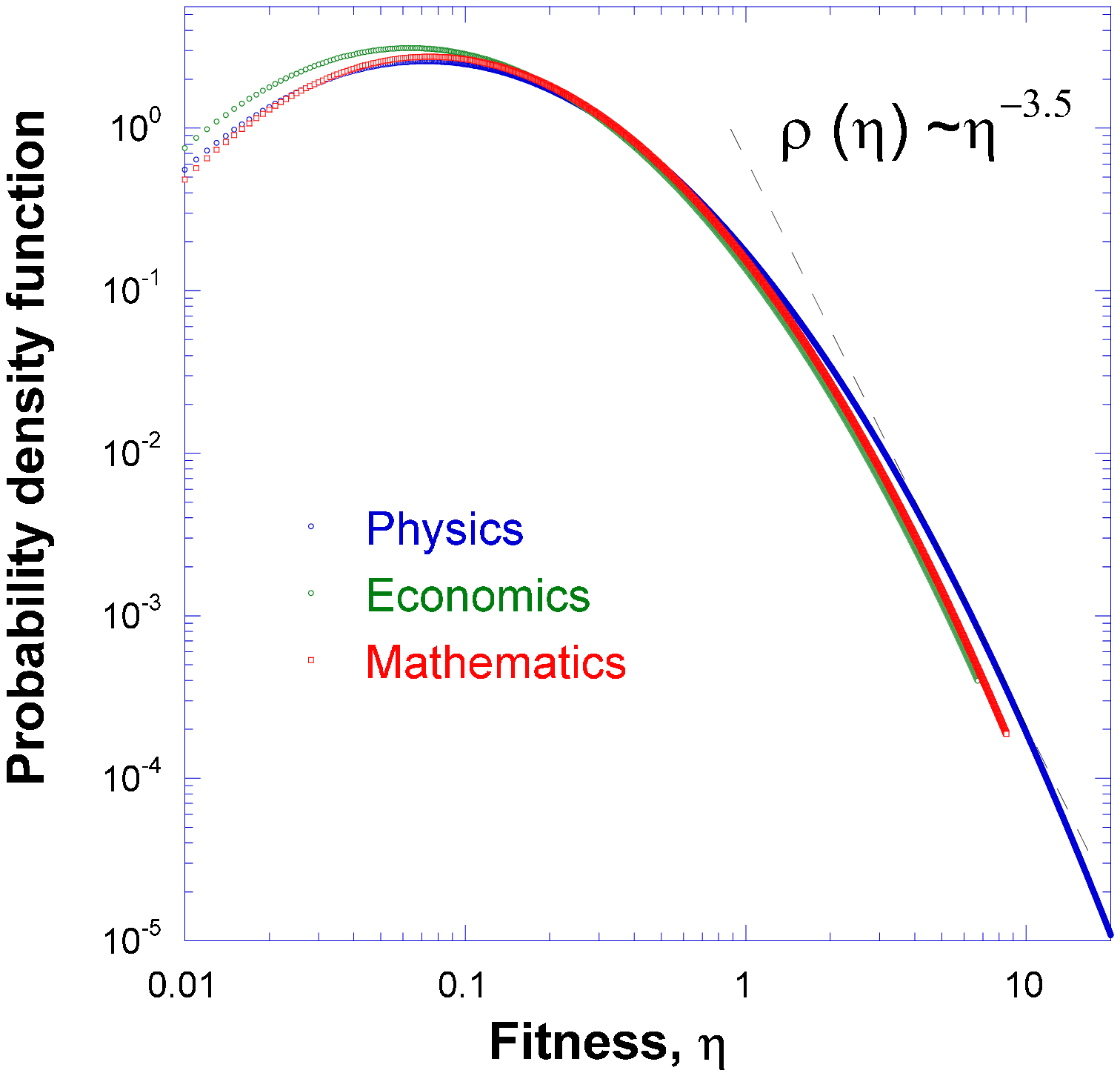}
\caption{ Continuous line shows fitness distributions $\rho(\eta)$ for Physics,  Economics, and Mathematics papers published in 1984. They  are fairly well approximated by   log-normals  with $\mu=-1.35,-1.54,-1.42$ and $\sigma=1.13, 1.10,1.08$ for Physics, Economics, and Mathematics papers, correspondingly.   The  mean is, correspondingly, $\eta_{0}=0.49, 0.39$ and $0.435$ while the maximum (mode) for all three distributions is $\eta_{max}\approx 0.07$.  The span of fitnesses is determined by the number of papers in each discipline. The dashed line shows the power-law approximation, $\rho(\eta)\propto\eta^{-3.5}$.
}
\label{fig:fitness}
\end{figure}

Figure \ref{fig:fitness}  shows that the fitness distributions $\rho(\eta)$ for all three disciplines are very similar and are represented by  the bell-shaped curves with the power-law tail, $\rho(\eta)\propto \eta^{-3.5}$.  A log-normal distribution   approximates these distributions fairly well.
\subsubsection{Indirect citations}
\begin{figure}[ht]
\includegraphics*[width=0.32\textwidth]{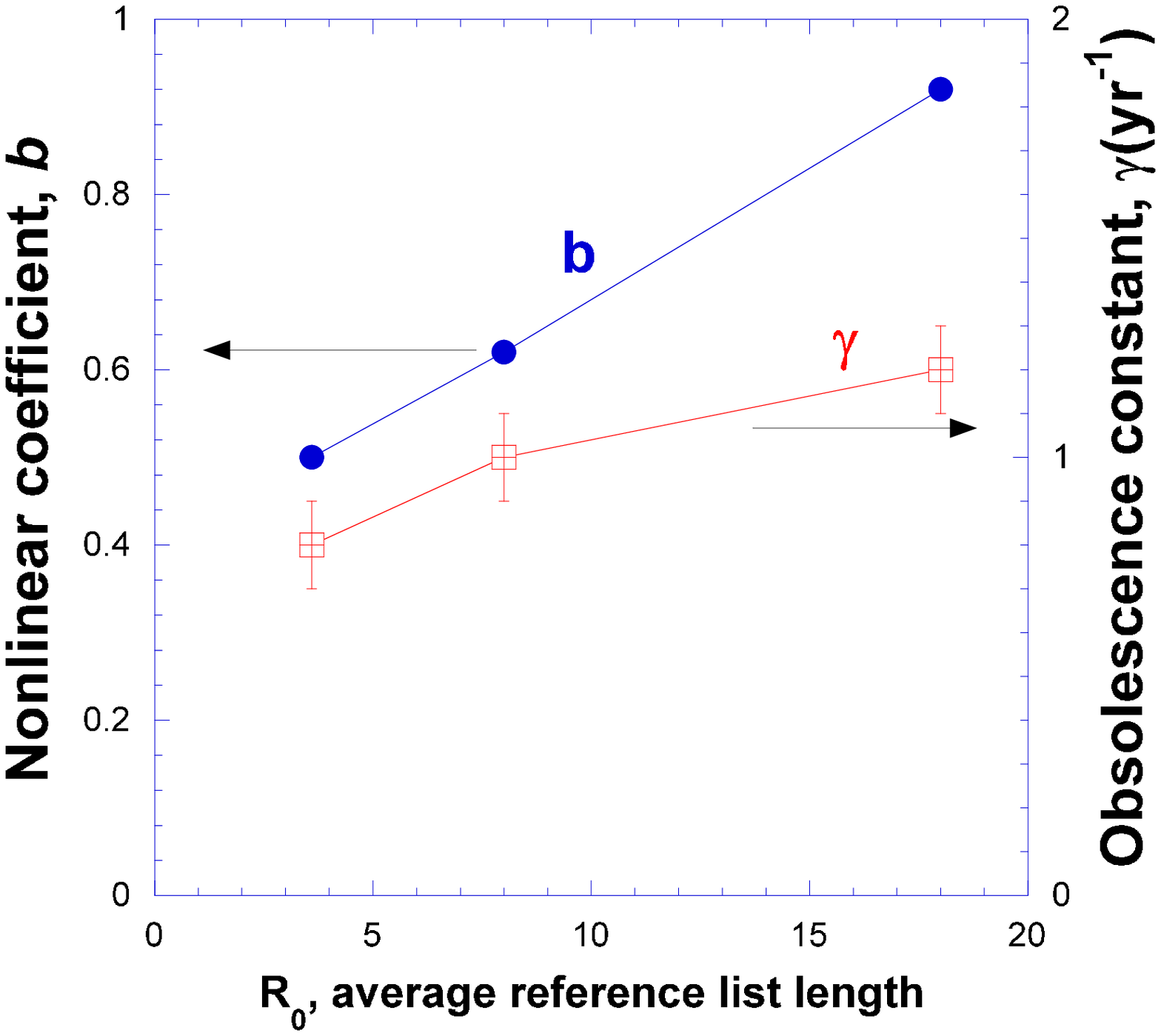}
\includegraphics*[width=0.32\textwidth]{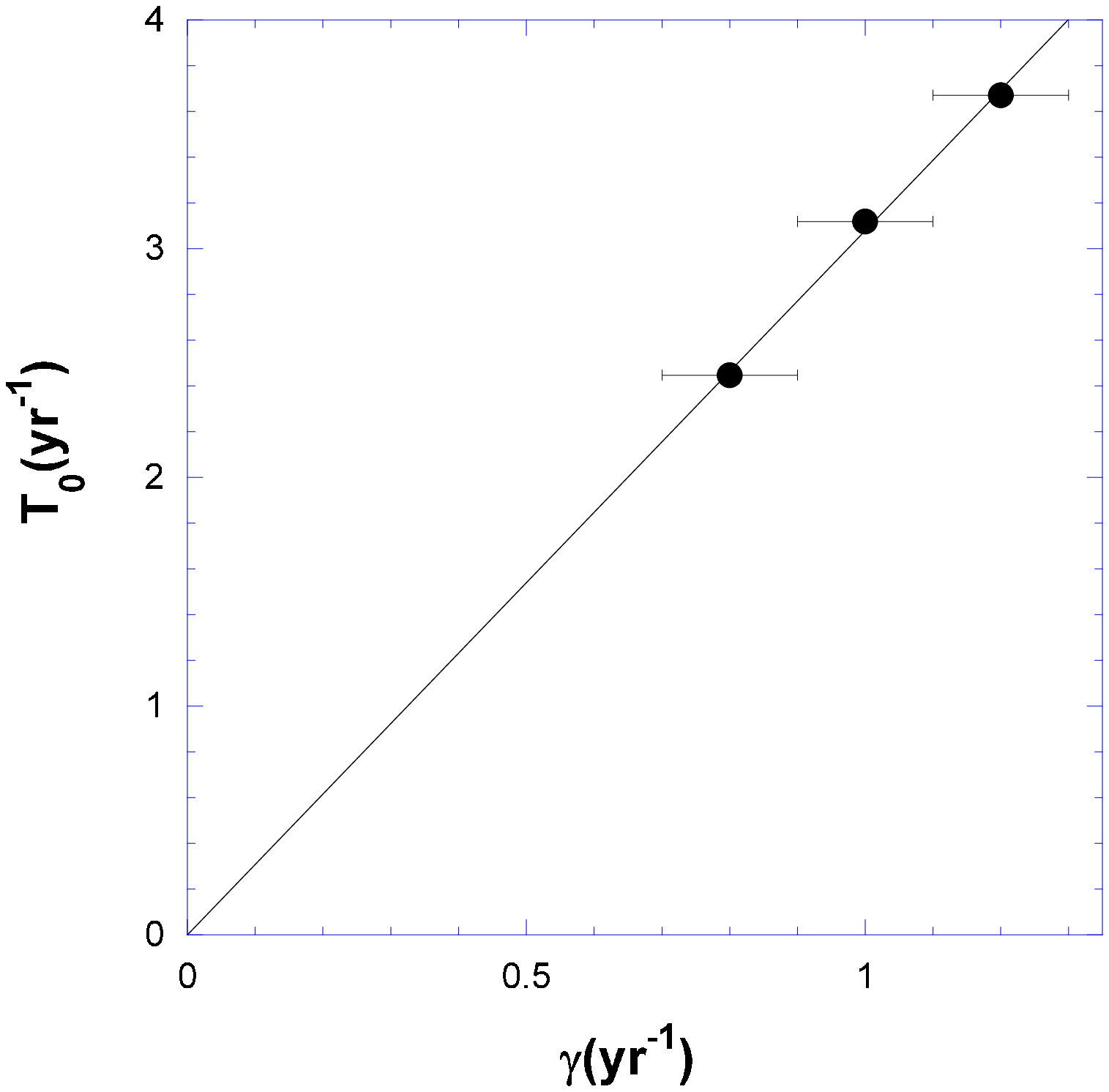}
\includegraphics*[width=0.32\textwidth]{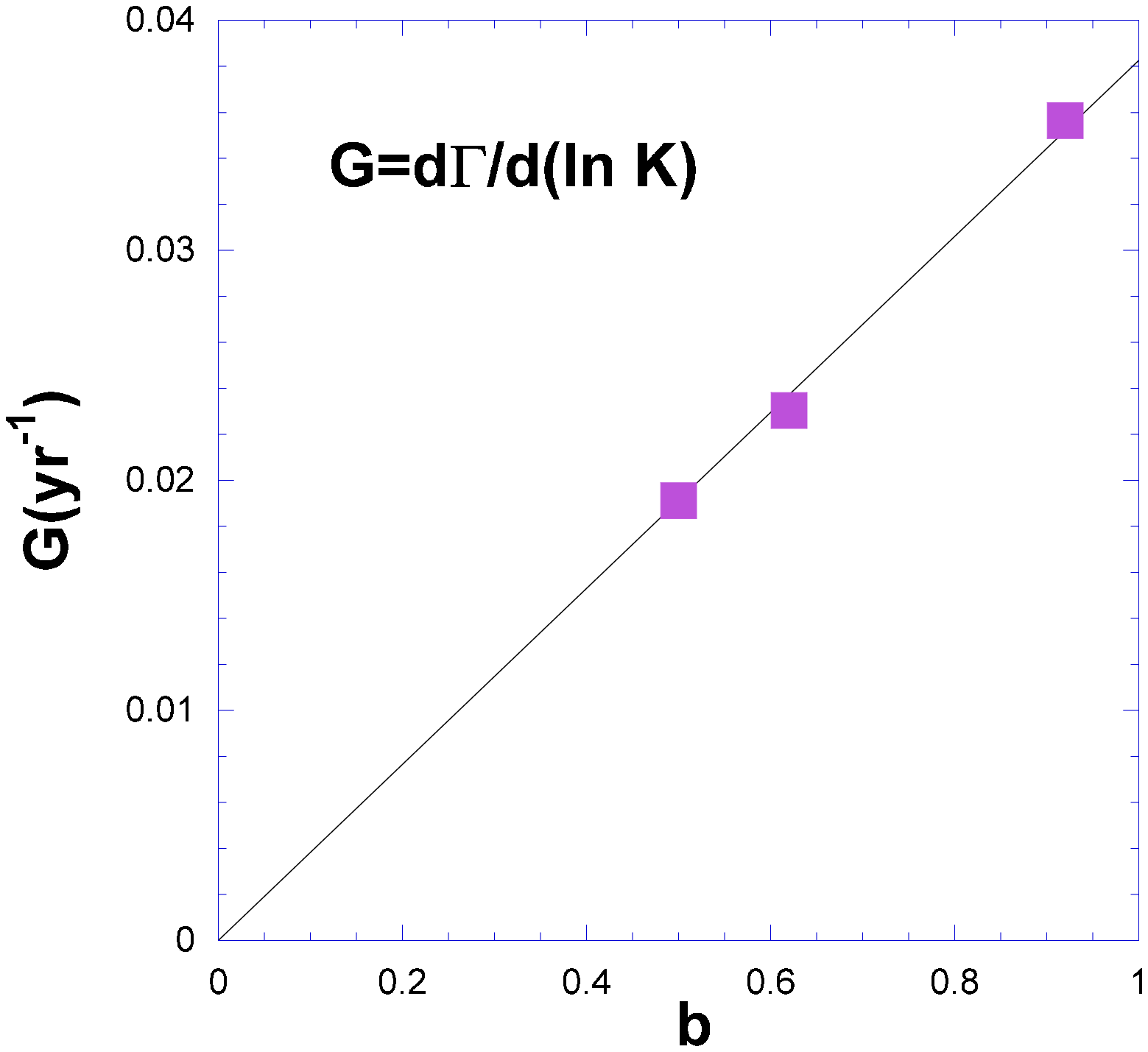}
\caption{The parameters   characterizing indirect citations for three disciplines. (a) Obsolescence rate $\gamma$ and nonlinear coefficient $b$ slowly grow with increasing  average reference list length $R_{0}$. Continuous lines are the guides to the eye. (b) $T_{0}$ is proportional to $\gamma$. (c) The parameter $G$ which characterizes citation lifetime (Eq. \ref{G}) is proportional to $b$, as expected.
}
\label{fig:indirect}
\end{figure}

Figure \ref{fig:indirect}  shows that the parameters  $\gamma, T_{0}$, and $b$, which characterize indirect citations, differ from discipline to discipline and systematically increase with the reference list length $R_{0}$.  With respect to the obsolescence constant $\gamma$ (Fig. \ref{fig:indirect}a), which is determined by the citing habits of researchers in each discipline, we do not perceive a clear reason why it should depend  on $R_{0}$. On another hand, since  $T_{0}$  and $\gamma$  have the same dimension  of yr$^{-1}$, and are the only such parameters characterizing the indirect citations, their proportionality (Fig. \ref{fig:indirect}b) is not unexpected. Regarding the nonlinear parameter $b$, it is determined by the connectivity of citation network and by the average community size \cite{Golosovsky2019}. Therefore, its dependence on $R_{0}$ (Fig. \ref{fig:indirect}a), which characterizes the average connectivity of citation network, is quite natural. Its proportionality to the coefficient $G$ in Eq. \ref{G} (Fig. \ref{fig:indirect}c) is also expected since $G$ characterizes the runaway behavior and the nonlinearity in the dynamic Eq. \ref{lambda} is the only source of runaways in our model.

In summary, our fitting procedure yields similar fitness distributions for all three disciplines. The parameters $\gamma$ and $T_{0}$  turn out to be discipline-specific, but their ratio is universal.  The nonlinear parameter $b$ is discipline-specific, the aging functions  $\tilde{A}(t)$ and the average reference lengths $R_{0}$ are  discipline-specific as well. Thus, while most parameters of citation dynamics are nearly identical for three different disciplines, there are few which are different.  As we will show further, scaling of the citation distributions  eliminates discipline-specific variability associated with $\tilde{A}(t)$ and  $R_{0}$, but not that associated with $b$ and $\gamma$. All this implies that citation distributions are nearly-universal and the deviations from  universality are related to indirect citations.

\section{Universality of  citation distributions explained}
We are now  in the position to assess the universality of citation distributions and the deviations therefrom. To this end, we consider Eq. \ref{lambda}, neglect its stochastic character, and  replace there $\lambda_{j}(\eta_{j},t)$, the  latent citation  rate of a paper $j$ in year $t$, by the actual citation rate $k_{j}(t)$.  Since  the exponent $e^{-\gamma(t-\tau)}$ decays with time much faster than the mean citation rate $m(t)$, we replace this slowly varying function by a constant   and lump together all prefactors in the integral  into one prefactor $q(K_{j})$. After these transformations Eq. \ref{lambda}  reduces to  nonlinear Fredholm integral equation of the second kind,
\begin{equation}
k_{j}(t)=\eta_{j}R_{0}\tilde{A}(t)+\int_{0}^{t}q(K_{j})e^{-\tilde{\gamma}(t-\tau)}k_{j}(\tau)d\tau,
\label{k}
\end{equation}
where $\tilde{\gamma}$ is a slightly modified obsolescence rate and $q(K_{j})=q_{0}(1+b\ln K_{j})$.  The nonlinearity comes just from this latter term.  In the first approximation,  we  replace it with a constant, $q$.  After such substitution, Eq. \ref{k}  is easily solved, yielding
\begin{equation}
k_{j}(t)=\eta_{j}R_{0}\left[\tilde{A}(t)+
q\int_{0}^{t}\tilde{A}(\tau)e^{-(\tilde{\gamma}-q)(t-\tau)}d\tau\right],
\label{SM-dynamics2}
\end{equation}
where the first addend in square brackets corresponds to direct citations and the second addend corresponds to indirect citations. Integration of  Eq. \ref{SM-dynamics2} yields
\begin{equation}
K_{j}(t)=\eta_{j}R_{0}B(t),
\label{SM-dynamics2a}
\end{equation}
where the time-dependent factor
\begin{equation}
B(t)=\int_{0}^{t}\tilde{A}(\tau)d\tau+
q\int_{0}^{t}dt_{1}\int_{0}^{t_{1}}\tilde{A}(\tau)e^{-(\tilde{\gamma}-q)(t_{1}-\tau)}d\tau
\label{SM-dynamics2b}
\end{equation}
is the same for all papers  in one discipline published in one year.

The mean number of citations is $M(t)=\overline{K_{j}(t)}$. Equation \ref{SM-dynamics2a} yields
\begin{equation}
M(t)=\eta_{0}R_{0}B(t),
\label{SM-dynamics2c}
\end{equation}
where $\eta_{0}=\int_{0}^{\infty}\eta\rho(\eta)d\eta$ is the mean fitness. Equations \ref  {SM-dynamics2a}, \ref{SM-dynamics2c}  yield  $\frac{K_{j}(t)}{M(t)}=\frac{\eta_{j}}{\eta_{0}}$. Thus,  scaled citation distribution is nothing else but  the reduced fitness distribution,
\begin{equation}
\rho\left(\frac{K(t)}{M(t)}\right)=\rho\left(\frac{\eta}{\eta_{0}}\right).
\label{scaled}
\end{equation}

Since  in the linear approximation, the time-dependent factor $B(t)$ and the discipline-dependent factor $R_{0}$ are scaled out, the only difference between  citation distributions for different collections of papers arises from the static factors which are all lumped in one parameter- fitness. Indeed, $\rho\left(\frac{\eta}{\eta_{0}}\right)$ in Eq. \ref{scaled} is nothing else but the reduced fitness distribution which, as Fig. \ref{fig:fitness} shows,  is close to a log-normal distribution with  almost the same width $\sigma$ and the same median $\mu=-\frac{\sigma^{2}}{2}$. Thus, the  near-universality of  citation distributions for different disciplines can be traced to the fact that the corresponding fitness distributions are log-normals with  the same width.

However, if  we compare between sets of papers  belonging to different journals, the scaling does not necessarily hold since the fitness distributions for the journal-based sets can be different.  Although these  distributions are subsets of the fitness distribution for the whole discipline,  the sampling performed  by each journal is not the same   due to  different acceptance criteria. In particular, for the journals shown in  Fig. \ref{fig:fitness}, the Science and the Physical Review Letters skim  the high-fitness tail of the fitness distribution for Physics while the Journal of Applied Physics samples uniformly  the whole distribution, and that is why these distributions do not scale.

Now we are in the position to explain the deviations from the universality of citation distributions. On the one hand, they are associated with slow growth of $q$ with $K$.  This $q(K)$ dependence introduces nonlinearity into Eq. \ref{SM-dynamics2}, its solution  can't be factorized. (Technically, this results in the modification of the function $B(t)$ which now acquires dependence on $\eta$.) Since the nonlinear parameter $b$ varies from discipline to discipline (Fig.\ref{fig:indirect}a) , the right-hand side of Eq. \ref{scaled} becomes discipline-specific  and the scaling given by Eq. \ref{scaled} breaks, namely, $\rho\left(\frac{K(t)}{M(t)}\right)\neq\rho\left(\frac{\eta}{\eta_{0}}\right)$.
However, the structure of Eq. \ref {T} suggests that $q(K)$ dependence is important mostly  for highly-cited papers, while for low-cited papers  $q(K)\approx const$. Thus, deviations of the scaled citation distributions from the universal distribution  shall be most prominent in their tails which consist of the highly-cited papers. Restricting  ourselves to one discipline, we note that  for the papers published in the same year, early citation distributions contain a very small number of highly-cited papers, hence  these distributions   scale (Fig. \ref{fig:physics-years}b), according to Eq. \ref{scaled}. However,  late citation distributions which contain many highly-cited papers, exhibit  significant deviations from  scaling (Fig. \ref{fig:physics-years}c). The same logic applies when we compare the sets of papers belonging to different disciplines and published in the same year. Since the fitness distributions for different disciplines are much more the same, citation distributions scale  as far  as the condition  $q(K)\approx const$ holds (Fig. \ref{fig:disciplines}b). This condition breaks for  highly-cited papers. Hence, the early  citation distributions for different disciplines scale while the late citation distributions do not scale. In both cases, the near-universality of citation distributions can be traced to universality of fitness distributions for the papers in different disciplines.


\section{Discussion}
 In our discussion of the presumed  universality of citation distributions, the shift of  focus from citations to fitnesses offered several advantages from the mathematical point of view.  This approach allows to decouple between deterministic and stochastic components of the citation process. In addition, it provides opportunity to analyze the uncited papers  since,  in our framework, these are not  zero-fitness but  low-fitness papers that, for purely statistical reasons, didn't get  their chance to be cited.  Another advantage of the fitness over citation distribution is that the former is continuous while the latter is discrete. Continuous distributions are  easier for analysis  and that is why we used the  log-normal fit in Fig. \ref{fig:fitness}.

 Although such fit has a long history in citation analysis, the fact that fitness distributions for three different disciplines follow  the  log-normal  dependence with  nearly the same $\sigma\sim 1.1$ is significant.  Notably, US patent citations are described by  the log-normal distribution with  the same $\sigma=1.1$  \cite{Clough2014}.  It should be noted that while log-normal distributions are ubiquitous in nature \cite{Limpert2001},   the distribution with  $\sigma\sim 1$ is  one of the narrowest observed. In fact, Ghadge et al. \cite{Ghadge2010} showed that such distribution  generates a citation network which is a borderline between the two classes- the gel-like network and the isolated clusters.

On another hand, the fact that fitness distributions for different disciplines have similar mean which is characterized by $\eta_{0}$ (Fig. \ref{fig:fitness}), is the consequence of nearly universal age distribution of references, as suggested by   Eq. \ref{rrr}.  Indeed, since $r(\tilde{\tau})$ is nearly universal \cite{Roth2012} and only weakly depends on time \cite{Sinatra2015}, the ratio $\frac{T_{0}}{\gamma}$ is also nearly universal (see Fig. \ref{fig:indirect}b), then $\eta_{0}$ shall be nearly universal as well.

In what follows we discuss more closely  the difference between citation distribution and fitness distribution.  In the framework of our recursive search model, the  information about a new paper propagates in the scientific community in  two ways: broadcasting (the authors find this paper after reading news, searching in the internet, reading the journals, etc.)   and word-of-mouth (finding this paper in the reference lists of another papers). The former way results in direct citations,  the latter way corresponds to indirect citations. These two ways of propagation are coupled: each direct citation gives rises to the cascade of indirect citations which can  become viral, in such a way that the paper becomes a runaway. While  direct citations of the paper are garnered in proportion to papers' fitness which captures its intrinsic quality and attributes, indirect citations depend on the structure of citation network and, in a sense, they gauge the papers' fame. While papers' fitness captures only its intrinsic attributes, the number of papers'  citations  combines together the papers' fitness and fame \cite{Simkin2013}. Since indirect citations are, in some sense,  consequences of direct citations, the papers' fitness  is the most important parameter that determines the number of citations that it acquires.  Our results imply that  fitness distributions  for different  disciplines are very similar. This is in contrast to  citation distributions  which are non-universal inasmuch as they are associated with propagation effects in the scientific network of each discipline.

The universality of fitness distributions for different disciplines  means that the proportion between the low and highly-cited papers for each of them is  more or less the same. This probably results from the same organizational and  hierarchical  structure of the research teams and institutions. Hierarchical structure is often associated with some kind of self-organized criticality  which results in the  power-law or Zipf-Mandelbrot distribution, $\Pi(\eta)\propto\frac{1}{(\eta+w)^{a}}$. The power-law tail of the fitness distribution  (Fig. \ref{fig:fitness}) may be an evidence of some kind of  the hierarchical structure of the corresponding scientific community  \cite{Leydesdorff2018}.

\begin{figure}[ht]
\includegraphics*[width=0.35
\textwidth]{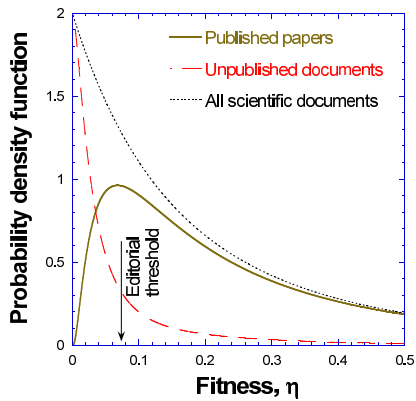}
\caption{Schematic drawing showing purported fitness distribution  for the set of all documents belonging to one discipline. While fitness distribution for all documents is expected to follow the power-law or  shifted power-law dependence,  $\Pi(\eta)\propto \frac{1}{(\eta+w)^{a}}$,  the corresponding distribution for the published papers rounds off at low fitnesses, in such a way that the whole distribution can be approximated by the log-normal dependence.  Thus, fitness distribution for the remaining documents, those that were not published in scientific journals, peaks at zero  fitness and then quickly decays.
}
\label{fig:fitness-scheme}
\end{figure}

However, if the fat-tail  fitness distribution is a consequence of the self-organized criticality, why does it deviate from the power-law dependence at low fitnesses and rounds off instead?   Below, we present a hand-waving explanation of this.  Consider the set of all possible scientific documents  belonging to one discipline such as papers, arXive submissions, drafts, reports, theses, etc.  According to the self-organized criticality hypothesis \cite{Leydesdorff2018},   we expect that  the  fitness distribution for the documents in this metaset follows the power-law dependence. Original research papers published in scientific journals represent only a subset of this metaset. In distinction to other documents,  they passed through the scrutinizing inspection of the  editors and reviewers, their fitness exceeds some threshold,  and that is why they have high publicity. The remaining documents, that were not screened and that do not have high publicity,   make  dominant contribution to the low-fitness  part of the fitness distribution. Thus, while the fitness distribution for the whole metaset most probably follows the power-law dependence, the fitness distribution for the subset of published papers rounds off at low fitnesses.  This low-fitness cutoff at $\eta_{max}\sim0.07$ is  the indication of the effectiveness of  the editors and reviewers. This value shall be compared to $\eta_{0}$, the mean fitness. Their ratio, $\frac{\eta_{0}}{\eta_{max}}$, indicate the acceptance threshold, namely, the lowest acceptable fitness with respect to the average one.  For three  disciplines that we studied here,  this threshold  is more or less the same and equals  to 14-18$\%$.   Roughly speaking,  this means that the editorial criteria for accepting the paper for publication are  such that the paper which, according to some subjective criteria of an editor, is five times less promising than the average one is still publishable, but the paper that is ten times less promising - is not. To our opinion, the fact, that this fitness threshold is more or less the same for three disciplines, is not surprising in view that disciplines are not disconnected, many studies are interdisciplinary, and interaction between the disciplines results in that the criteria for editorial acceptance become aligned.

If different disciplines exhibit almost the same fitness distribution, why the distributions for different journals are not the same? To our opinion, the reason is that each scientific discipline can be viewed as an almost closed system which undergoes self-organization, while a single journal is not a closed system. Indeed, the researchers in one discipline predominantly cite other studies belonging to the same discipline, while the journals in one discipline are not that introspective and heavily cite one another.

\section{Summary}
Citation distributions are determined by the fitness distribution and by the citation dynamics of papers. While citation distributions differ from discipline to discipline, the fitness distribution is nearly universal, at least for Physics, Economics, and Mathematics, and it is well-approximated by the log-normal distribution with $\sigma\sim 1.1$. This fat-tailed distribution probably reveals some facet of science as a  self-organizing system.

Universality of citation distributions holds for collections of papers belonging to one discipline, published in one year, and measured several years after publication. The underlying reason for this universality is the nearly universal fitness distribution.  Universality of citation distributions does not hold when one compares collections of papers many years after publication (deviations are associated with the discipline-specific citation dynamics) or for different journals (deviations are associated with the journal-specific fitness distribution).

\begin{acknowledgments}
 I am grateful to Sorin Solomon for fruitful discussions, to Magda Fontana for  stimulating discussions and for the  assessment of the Economics journals,  and to Yakov Varshavsky for the  assessment of the Mathematics journals.
\end{acknowledgments}
\bibliography{reference_master_2019_new}

\begin{thebibliography}{35}%
\makeatletter
\providecommand \@ifxundefined [1]{%
 \@ifx{#1\undefined}
}%
\providecommand \@ifnum [1]{%
 \ifnum #1\expandafter \@firstoftwo
 \else \expandafter \@secondoftwo
 \fi
}%
\providecommand \@ifx [1]{%
 \ifx #1\expandafter \@firstoftwo
 \else \expandafter \@secondoftwo
 \fi
}%
\providecommand \natexlab [1]{#1}%
\providecommand \enquote  [1]{``#1''}%
\providecommand \bibnamefont  [1]{#1}%
\providecommand \bibfnamefont [1]{#1}%
\providecommand \citenamefont [1]{#1}%
\providecommand \href@noop [0]{\@secondoftwo}%
\providecommand \href [0]{\begingroup \@sanitize@url \@href}%
\providecommand \@href[1]{\@@startlink{#1}\@@href}%
\providecommand \@@href[1]{\endgroup#1\@@endlink}%
\providecommand \@sanitize@url [0]{\catcode `\\12\catcode `\$12\catcode
  `\&12\catcode `\#12\catcode `\^12\catcode `\_12\catcode `\%12\relax}%
\providecommand \@@startlink[1]{}%
\providecommand \@@endlink[0]{}%
\providecommand \url  [0]{\begingroup\@sanitize@url \@url }%
\providecommand \@url [1]{\endgroup\@href {#1}{\urlprefix }}%
\providecommand \urlprefix  [0]{URL }%
\providecommand \Eprint [0]{\href }%
\providecommand \doibase [0]{http://dx.doi.org/}%
\providecommand \selectlanguage [0]{\@gobble}%
\providecommand \bibinfo  [0]{\@secondoftwo}%
\providecommand \bibfield  [0]{\@secondoftwo}%
\providecommand \translation [1]{[#1]}%
\providecommand \BibitemOpen [0]{}%
\providecommand \bibitemStop [0]{}%
\providecommand \bibitemNoStop [0]{.\EOS\space}%
\providecommand \EOS [0]{\spacefactor3000\relax}%
\providecommand \BibitemShut  [1]{\csname bibitem#1\endcsname}%
\let\auto@bib@innerbib\@empty
\bibitem [{\citenamefont {Barabasi}(2015)}]{Barabasi2015}%
  \BibitemOpen
  \bibfield  {author} {\bibinfo {author} {\bibfnamefont {A.-L.}\ \bibnamefont
  {Barabasi}},\ }\href
  {http://www.ebook.de/de/product/24312547/albert_laszlo_barabasi_network_science.html}
  {\emph {\bibinfo {title} {Network Science}}}\ (\bibinfo  {publisher}
  {Cambridge University Press},\ \bibinfo {year} {2015})\BibitemShut {NoStop}%
\bibitem [{\citenamefont {Price}(1976)}]{Price1976}%
  \BibitemOpen
  \bibfield  {author} {\bibinfo {author} {\bibfnamefont {D.~D.~S.}\
  \bibnamefont {Price}},\ }\href {\doibase 10.1002/asi.4630270505} {\bibfield
  {journal} {\bibinfo  {journal} {J. Am. Soc. Inf. Sci.}\ }\textbf {\bibinfo
  {volume} {27}},\ \bibinfo {pages} {292} (\bibinfo {year} {1976})}\BibitemShut
  {NoStop}%
\bibitem [{\citenamefont {Seglen}(1992)}]{Seglen1992}%
  \BibitemOpen
  \bibfield  {author} {\bibinfo {author} {\bibfnamefont {P.~O.}\ \bibnamefont
  {Seglen}},\ }\href {\doibase
  10.1002/(sici)1097-4571(199210)43:9<628::aid-asi5>3.0.co;2-0} {\bibfield
  {journal} {\bibinfo  {journal} {J. Am. Soc. Inf. Sci.}\ }\textbf {\bibinfo
  {volume} {43}},\ \bibinfo {pages} {628} (\bibinfo {year} {1992})}\BibitemShut
  {NoStop}%
\bibitem [{\citenamefont {Radicchi}\ \emph {et~al.}(2008)\citenamefont
  {Radicchi}, \citenamefont {Fortunato},\ and\ \citenamefont
  {Castellano}}]{Radicchi2008}%
  \BibitemOpen
  \bibfield  {author} {\bibinfo {author} {\bibfnamefont {F.}~\bibnamefont
  {Radicchi}}, \bibinfo {author} {\bibfnamefont {S.}~\bibnamefont {Fortunato}},
  \ and\ \bibinfo {author} {\bibfnamefont {C.}~\bibnamefont {Castellano}},\
  }\href {\doibase 10.1073/pnas.0806977105} {\bibfield  {journal} {\bibinfo
  {journal} {Proceedings of the National Academy of Sciences}\ }\textbf
  {\bibinfo {volume} {105}},\ \bibinfo {pages} {17268} (\bibinfo {year}
  {2008})}\BibitemShut {NoStop}%
\bibitem [{\citenamefont {Barzel}\ and\ \citenamefont
  {Barabasi}(2013)}]{Barzel2013}%
  \BibitemOpen
  \bibfield  {author} {\bibinfo {author} {\bibfnamefont {B.}~\bibnamefont
  {Barzel}}\ and\ \bibinfo {author} {\bibfnamefont {A.-L.}\ \bibnamefont
  {Barabasi}},\ }\href {\doibase 10.1038/nbt.2601} {\bibfield  {journal}
  {\bibinfo  {journal} {Nature Biotechnology}\ }\textbf {\bibinfo {volume}
  {31}},\ \bibinfo {pages} {720} (\bibinfo {year} {2013})}\BibitemShut
  {NoStop}%
\bibitem [{\citenamefont {Gao}\ \emph {et~al.}(2016)\citenamefont {Gao},
  \citenamefont {Barzel},\ and\ \citenamefont {Barabasi}}]{Gao2016}%
  \BibitemOpen
  \bibfield  {author} {\bibinfo {author} {\bibfnamefont {J.}~\bibnamefont
  {Gao}}, \bibinfo {author} {\bibfnamefont {B.}~\bibnamefont {Barzel}}, \ and\
  \bibinfo {author} {\bibfnamefont {A.-L.}\ \bibnamefont {Barabasi}},\ }\href
  {https://doi.org/10.1038/nature16948} {\bibfield  {journal} {\bibinfo
  {journal} {Nature}\ }\textbf {\bibinfo {volume} {530}},\ \bibinfo {pages}
  {307} (\bibinfo {year} {2016})}\BibitemShut {NoStop}%
\bibitem [{\citenamefont {Candia}\ \emph {et~al.}(2019)\citenamefont {Candia},
  \citenamefont {Jara-Figueroa}, \citenamefont {Rodriguez-Sickert},
  \citenamefont {Barabasi},\ and\ \citenamefont {Hidalgo}}]{Candia2019}%
  \BibitemOpen
  \bibfield  {author} {\bibinfo {author} {\bibfnamefont {C.}~\bibnamefont
  {Candia}}, \bibinfo {author} {\bibfnamefont {C.}~\bibnamefont
  {Jara-Figueroa}}, \bibinfo {author} {\bibfnamefont {C.}~\bibnamefont
  {Rodriguez-Sickert}}, \bibinfo {author} {\bibfnamefont {A.-L.}\ \bibnamefont
  {Barabasi}}, \ and\ \bibinfo {author} {\bibfnamefont {C.~A.}\ \bibnamefont
  {Hidalgo}},\ }\href {https://doi.org/10.1038/s41562-018-0474-5} {\bibfield
  {journal} {\bibinfo  {journal} {Nature Human Behaviour}\ }\textbf {\bibinfo
  {volume} {3}},\ \bibinfo {pages} {82} (\bibinfo {year} {2019})}\BibitemShut
  {NoStop}%
\bibitem [{\citenamefont {Fortunato}\ \emph {et~al.}(2018)\citenamefont
  {Fortunato}, \citenamefont {Bergstrom}, \citenamefont {Borner}, \citenamefont
  {Evans}, \citenamefont {Helbing}, \citenamefont {Milojevic}, \citenamefont
  {Petersen}, \citenamefont {Radicchi}, \citenamefont {Sinatra}, \citenamefont
  {Uzzi}, \citenamefont {Vespignani}, \citenamefont {Waltman}, \citenamefont
  {Wang},\ and\ \citenamefont {Barabasi}}]{Fortunato2018}%
  \BibitemOpen
  \bibfield  {author} {\bibinfo {author} {\bibfnamefont {S.}~\bibnamefont
  {Fortunato}}, \bibinfo {author} {\bibfnamefont {C.~T.}\ \bibnamefont
  {Bergstrom}}, \bibinfo {author} {\bibfnamefont {K.}~\bibnamefont {Borner}},
  \bibinfo {author} {\bibfnamefont {J.~A.}\ \bibnamefont {Evans}}, \bibinfo
  {author} {\bibfnamefont {D.}~\bibnamefont {Helbing}}, \bibinfo {author}
  {\bibfnamefont {S.}~\bibnamefont {Milojevic}}, \bibinfo {author}
  {\bibfnamefont {A.~M.}\ \bibnamefont {Petersen}}, \bibinfo {author}
  {\bibfnamefont {F.}~\bibnamefont {Radicchi}}, \bibinfo {author}
  {\bibfnamefont {R.}~\bibnamefont {Sinatra}}, \bibinfo {author} {\bibfnamefont
  {B.}~\bibnamefont {Uzzi}}, \bibinfo {author} {\bibfnamefont {A.}~\bibnamefont
  {Vespignani}}, \bibinfo {author} {\bibfnamefont {L.}~\bibnamefont {Waltman}},
  \bibinfo {author} {\bibfnamefont {D.}~\bibnamefont {Wang}}, \ and\ \bibinfo
  {author} {\bibfnamefont {A.-L.}\ \bibnamefont {Barabasi}},\ }\href
  {http://science.sciencemag.org/content/359/6379/eaao0185.abstract} {\bibfield
   {journal} {\bibinfo  {journal} {Science}\ }\textbf {\bibinfo {volume}
  {359}},\ \bibinfo {pages} {eaao0185} (\bibinfo {year} {2018})}\BibitemShut
  {NoStop}%
\bibitem [{\citenamefont {Kuhn}(1970)}]{Kuhn1970}%
  \BibitemOpen
  \bibfield  {author} {\bibinfo {author} {\bibfnamefont {T.~S.}\ \bibnamefont
  {Kuhn}},\ }\href@noop {} {\emph {\bibinfo {title} {The structure of
  scientific revolutions}}}\ (\bibinfo  {publisher} {University of Chicago
  Press Ltd},\ \bibinfo {year} {1970})\BibitemShut {NoStop}%
\bibitem [{\citenamefont {Bornmann}\ and\ \citenamefont
  {Daniel}(2009)}]{Bornmann2009}%
  \BibitemOpen
  \bibfield  {author} {\bibinfo {author} {\bibfnamefont {L.}~\bibnamefont
  {Bornmann}}\ and\ \bibinfo {author} {\bibfnamefont {H.-D.}\ \bibnamefont
  {Daniel}},\ }\href {\doibase 10.1002/asi.21076} {\bibfield  {journal}
  {\bibinfo  {journal} {J. Am. Soc. Inf. Sci.}\ }\textbf {\bibinfo {volume}
  {60}},\ \bibinfo {pages} {1664} (\bibinfo {year} {2009})}\BibitemShut
  {NoStop}%
\bibitem [{\citenamefont {Waltman}\ \emph {et~al.}(2011)\citenamefont
  {Waltman}, \citenamefont {van Eck},\ and\ \citenamefont {van
  Raan}}]{Waltman2011}%
  \BibitemOpen
  \bibfield  {author} {\bibinfo {author} {\bibfnamefont {L.}~\bibnamefont
  {Waltman}}, \bibinfo {author} {\bibfnamefont {N.~J.}\ \bibnamefont {van
  Eck}}, \ and\ \bibinfo {author} {\bibfnamefont {A.~F.~J.}\ \bibnamefont {van
  Raan}},\ }\href {\doibase 10.1002/asi.21671} {\bibfield  {journal} {\bibinfo
  {journal} {J. Am. Soc. Inf. Sci.}\ }\textbf {\bibinfo {volume} {63}},\
  \bibinfo {pages} {72} (\bibinfo {year} {2011})}\BibitemShut {NoStop}%
\bibitem [{\citenamefont {Evans}\ \emph {et~al.}(2012)\citenamefont {Evans},
  \citenamefont {Hopkins},\ and\ \citenamefont {Kaube}}]{Evans2012}%
  \BibitemOpen
  \bibfield  {author} {\bibinfo {author} {\bibfnamefont {T.~S.}\ \bibnamefont
  {Evans}}, \bibinfo {author} {\bibfnamefont {N.}~\bibnamefont {Hopkins}}, \
  and\ \bibinfo {author} {\bibfnamefont {B.~S.}\ \bibnamefont {Kaube}},\ }\href
  {\doibase 10.1007/s11192-012-0694-9} {\bibfield  {journal} {\bibinfo
  {journal} {Scientometrics}\ }\textbf {\bibinfo {volume} {93}},\ \bibinfo
  {pages} {473} (\bibinfo {year} {2012})}\BibitemShut {NoStop}%
\bibitem [{\citenamefont {Chatterjee}\ \emph {et~al.}(2016)\citenamefont
  {Chatterjee}, \citenamefont {Ghosh},\ and\ \citenamefont
  {Chakrabarti}}]{Chatterjee2016}%
  \BibitemOpen
  \bibfield  {author} {\bibinfo {author} {\bibfnamefont {A.}~\bibnamefont
  {Chatterjee}}, \bibinfo {author} {\bibfnamefont {A.}~\bibnamefont {Ghosh}}, \
  and\ \bibinfo {author} {\bibfnamefont {B.~K.}\ \bibnamefont {Chakrabarti}},\
  }\href {\doibase 10.1371/journal.pone.0146762} {\bibfield  {journal}
  {\bibinfo  {journal} {PLOS ONE}\ }\textbf {\bibinfo {volume} {11}},\ \bibinfo
  {pages} {e0146762} (\bibinfo {year} {2016})},\ \Eprint
  {http://arxiv.org/abs/1409.8029} {1409.8029} \BibitemShut {NoStop}%
\bibitem [{\citenamefont {D'Angelo}\ and\ \citenamefont
  {Di~Russo}(2019)}]{DAngelo2019}%
  \BibitemOpen
  \bibfield  {author} {\bibinfo {author} {\bibfnamefont {C.~A.}\ \bibnamefont
  {D'Angelo}}\ and\ \bibinfo {author} {\bibfnamefont {S.}~\bibnamefont
  {Di~Russo}},\ }\href
  {http://www.sciencedirect.com/science/article/pii/S1751157718302876}
  {\bibfield  {journal} {\bibinfo  {journal} {Journal of Informetrics}\
  }\textbf {\bibinfo {volume} {13}},\ \bibinfo {pages} {726} (\bibinfo {year}
  {2019})}\BibitemShut {NoStop}%
\bibitem [{\citenamefont {Radicchi}\ and\ \citenamefont
  {Castellano}(2011)}]{Radicchi2011}%
  \BibitemOpen
  \bibfield  {author} {\bibinfo {author} {\bibfnamefont {F.}~\bibnamefont
  {Radicchi}}\ and\ \bibinfo {author} {\bibfnamefont {C.}~\bibnamefont
  {Castellano}},\ }\href {\doibase 10.1103/physreve.83.046116} {\bibfield
  {journal} {\bibinfo  {journal} {Physical Review E}\ }\textbf {\bibinfo
  {volume} {83}},\ \bibinfo {pages} {046116} (\bibinfo {year}
  {2011})}\BibitemShut {NoStop}%
\bibitem [{\citenamefont {Radicchi}\ and\ \citenamefont
  {Castellano}(2012)}]{Radicchi2012}%
  \BibitemOpen
  \bibfield  {author} {\bibinfo {author} {\bibfnamefont {F.}~\bibnamefont
  {Radicchi}}\ and\ \bibinfo {author} {\bibfnamefont {C.}~\bibnamefont
  {Castellano}},\ }\href {\doibase 10.1371/journal.pone.0033833} {\bibfield
  {journal} {\bibinfo  {journal} {{PLoS} {ONE}}\ }\textbf {\bibinfo {volume}
  {7}},\ \bibinfo {pages} {e33833} (\bibinfo {year} {2012})}\BibitemShut
  {NoStop}%
\bibitem [{\citenamefont {Stringer}\ \emph {et~al.}(2008)\citenamefont
  {Stringer}, \citenamefont {Sales-Pardo},\ and\ \citenamefont
  {Amaral}}]{Stringer2008}%
  \BibitemOpen
  \bibfield  {author} {\bibinfo {author} {\bibfnamefont {M.~J.}\ \bibnamefont
  {Stringer}}, \bibinfo {author} {\bibfnamefont {M.}~\bibnamefont
  {Sales-Pardo}}, \ and\ \bibinfo {author} {\bibfnamefont {L.~A.~N.}\
  \bibnamefont {Amaral}},\ }\href {\doibase 10.1371/journal.pone.0001683}
  {\bibfield  {journal} {\bibinfo  {journal} {{PLoS} {ONE}}\ }\textbf {\bibinfo
  {volume} {3}},\ \bibinfo {pages} {e1683} (\bibinfo {year}
  {2008})}\BibitemShut {NoStop}%
\bibitem [{\citenamefont {Thelwall}(2016{\natexlab{a}})}]{Thelwall2016a}%
  \BibitemOpen
  \bibfield  {author} {\bibinfo {author} {\bibfnamefont {M.}~\bibnamefont
  {Thelwall}},\ }\href
  {http://www.sciencedirect.com/science/article/pii/S1751157716300517}
  {\bibfield  {journal} {\bibinfo  {journal} {Journal of Informetrics}\
  }\textbf {\bibinfo {volume} {10}},\ \bibinfo {pages} {863} (\bibinfo {year}
  {2016}{\natexlab{a}})}\BibitemShut {NoStop}%
\bibitem [{\citenamefont {Golosovsky}\ and\ \citenamefont
  {Solomon}(2017)}]{Golosovsky2017}%
  \BibitemOpen
  \bibfield  {author} {\bibinfo {author} {\bibfnamefont {M.}~\bibnamefont
  {Golosovsky}}\ and\ \bibinfo {author} {\bibfnamefont {S.}~\bibnamefont
  {Solomon}},\ }\href {\doibase 10.1103/physreve.95.012324} {\bibfield
  {journal} {\bibinfo  {journal} {Physical Review E}\ }\textbf {\bibinfo
  {volume} {95}},\ \bibinfo {pages} {012324} (\bibinfo {year}
  {2017})}\BibitemShut {NoStop}%
\bibitem [{\citenamefont {Golosovsky}(2019)}]{Golosovsky2019}%
  \BibitemOpen
  \bibfield  {author} {\bibinfo {author} {\bibfnamefont {M.}~\bibnamefont
  {Golosovsky}},\ }\href@noop {} {\emph {\bibinfo {title} {Citation Analysis
  and Dynamics of Citation Networks}}}\ (\bibinfo  {publisher} {Springer
  International Publishing},\ \bibinfo {year} {2019})\BibitemShut {NoStop}%
\bibitem [{\citenamefont {Simkin}\ and\ \citenamefont
  {Roychowdhury}(2007)}]{Simkin2007}%
  \BibitemOpen
  \bibfield  {author} {\bibinfo {author} {\bibfnamefont {M.~V.}\ \bibnamefont
  {Simkin}}\ and\ \bibinfo {author} {\bibfnamefont {V.~P.}\ \bibnamefont
  {Roychowdhury}},\ }\href {\doibase 10.1002/asi.20653} {\bibfield  {journal}
  {\bibinfo  {journal} {J. Am. Soc. Inf. Sci.}\ }\textbf {\bibinfo {volume}
  {58}},\ \bibinfo {pages} {1661} (\bibinfo {year} {2007})}\BibitemShut
  {NoStop}%
\bibitem [{\citenamefont {Thelwall}(2016{\natexlab{b}})}]{Thelwall2016b}%
  \BibitemOpen
  \bibfield  {author} {\bibinfo {author} {\bibfnamefont {M.}~\bibnamefont
  {Thelwall}},\ }\href
  {http://www.sciencedirect.com/science/article/pii/S1751157716300153}
  {\bibfield  {journal} {\bibinfo  {journal} {Journal of Informetrics}\
  }\textbf {\bibinfo {volume} {10}},\ \bibinfo {pages} {622} (\bibinfo {year}
  {2016}{\natexlab{b}})}\BibitemShut {NoStop}%
\bibitem [{\citenamefont {Bianconi}\ and\ \citenamefont
  {Barabasi}(2001)}]{Bianconi2001}%
  \BibitemOpen
  \bibfield  {author} {\bibinfo {author} {\bibfnamefont {G.}~\bibnamefont
  {Bianconi}}\ and\ \bibinfo {author} {\bibfnamefont {A.-L.}\ \bibnamefont
  {Barabasi}},\ }\href {\doibase 10.1103/PhysRevLett.86.5632} {\bibfield
  {journal} {\bibinfo  {journal} {Phys. Rev. Lett.}\ }\textbf {\bibinfo
  {volume} {86}},\ \bibinfo {pages} {5632} (\bibinfo {year}
  {2001})}\BibitemShut {NoStop}%
\bibitem [{\citenamefont {Caldarelli}\ \emph {et~al.}(2002)\citenamefont
  {Caldarelli}, \citenamefont {Capocci}, \citenamefont {DeLosRios},\ and\
  \citenamefont {Mu{\~{n}}oz}}]{Caldarelli2002}%
  \BibitemOpen
  \bibfield  {author} {\bibinfo {author} {\bibfnamefont {G.}~\bibnamefont
  {Caldarelli}}, \bibinfo {author} {\bibfnamefont {A.}~\bibnamefont {Capocci}},
  \bibinfo {author} {\bibfnamefont {P.}~\bibnamefont {DeLosRios}}, \ and\
  \bibinfo {author} {\bibfnamefont {M.~A.}\ \bibnamefont {Mu{\~{n}}oz}},\
  }\href {\doibase 10.1103/physrevlett.89.258702} {\bibfield  {journal}
  {\bibinfo  {journal} {Phys. Rev. Lett.}\ }\textbf {\bibinfo {volume} {89}},\
  \bibinfo {pages} {258702} (\bibinfo {year} {2002})}\BibitemShut {NoStop}%
\bibitem [{\citenamefont {Nakamoto}(1988)}]{Nakamoto1988}%
  \BibitemOpen
  \bibfield  {author} {\bibinfo {author} {\bibfnamefont {H.}~\bibnamefont
  {Nakamoto}},\ }\href {http://hdl.handle.net/1942/837} {\bibfield  {journal}
  {\bibinfo  {journal} {Informetrics}\ }\textbf {\bibinfo {volume} {87/88}},\
  \bibinfo {pages} {157} (\bibinfo {year} {1988})}\BibitemShut {NoStop}%
\bibitem [{\citenamefont {Glanzel}(2004)}]{Glanzel2004}%
  \BibitemOpen
  \bibfield  {author} {\bibinfo {author} {\bibfnamefont {W.}~\bibnamefont
  {Glanzel}},\ }\href {\doibase 10.1023/B:SCIE.0000034391.06240.2a} {\bibfield
  {journal} {\bibinfo  {journal} {Scientometrics}\ }\textbf {\bibinfo {volume}
  {60}},\ \bibinfo {pages} {511} (\bibinfo {year} {2004})}\BibitemShut
  {NoStop}%
\bibitem [{\citenamefont {Roth}\ \emph {et~al.}(2012)\citenamefont {Roth},
  \citenamefont {Wu},\ and\ \citenamefont {Lozano}}]{Roth2012}%
  \BibitemOpen
  \bibfield  {author} {\bibinfo {author} {\bibfnamefont {C.}~\bibnamefont
  {Roth}}, \bibinfo {author} {\bibfnamefont {J.}~\bibnamefont {Wu}}, \ and\
  \bibinfo {author} {\bibfnamefont {S.}~\bibnamefont {Lozano}},\ }\href
  {\doibase http://dx.doi.org/10.1016/j.joi.2011.08.005} {\bibfield  {journal}
  {\bibinfo  {journal} {Journal of Informetrics}\ }\textbf {\bibinfo {volume}
  {6}},\ \bibinfo {pages} {111 } (\bibinfo {year} {2012})}\BibitemShut
  {NoStop}%
\bibitem [{\citenamefont {Golosovsky}(2017)}]{Golosovsky2017a}%
  \BibitemOpen
  \bibfield  {author} {\bibinfo {author} {\bibfnamefont {M.}~\bibnamefont
  {Golosovsky}},\ }\href@noop {} {\bibfield  {journal} {\bibinfo  {journal}
  {Physical Review E}\ }\textbf {\bibinfo {volume} {96}},\ \bibinfo {pages}
  {032306} (\bibinfo {year} {2017})}\BibitemShut {NoStop}%
\bibitem [{\citenamefont {Golosovsky}\ and\ \citenamefont
  {Solomon}(2012)}]{Golosovsky2012a}%
  \BibitemOpen
  \bibfield  {author} {\bibinfo {author} {\bibfnamefont {M.}~\bibnamefont
  {Golosovsky}}\ and\ \bibinfo {author} {\bibfnamefont {S.}~\bibnamefont
  {Solomon}},\ }\href {\doibase 10.1140/epjst/e2012-01576-4} {\bibfield
  {journal} {\bibinfo  {journal} {The European Physical Journal}\ }\textbf
  {\bibinfo {volume} {205}},\ \bibinfo {pages} {303} (\bibinfo {year}
  {2012})}\BibitemShut {NoStop}%
\bibitem [{\citenamefont {Clough}\ \emph {et~al.}(2014)\citenamefont {Clough},
  \citenamefont {Gollings}, \citenamefont {Loach},\ and\ \citenamefont
  {Evans}}]{Clough2014}%
  \BibitemOpen
  \bibfield  {author} {\bibinfo {author} {\bibfnamefont {J.~R.}\ \bibnamefont
  {Clough}}, \bibinfo {author} {\bibfnamefont {J.}~\bibnamefont {Gollings}},
  \bibinfo {author} {\bibfnamefont {T.~V.}\ \bibnamefont {Loach}}, \ and\
  \bibinfo {author} {\bibfnamefont {T.~S.}\ \bibnamefont {Evans}},\ }\href
  {\doibase 10.1093/comnet/cnu039} {\bibfield  {journal} {\bibinfo  {journal}
  {Journal of Complex Networks}\ }\textbf {\bibinfo {volume} {3}},\ \bibinfo
  {pages} {189} (\bibinfo {year} {2014})}\BibitemShut {NoStop}%
\bibitem [{\citenamefont {Limpert}\ \emph {et~al.}(2001)\citenamefont
  {Limpert}, \citenamefont {Stahel},\ and\ \citenamefont {Abbt}}]{Limpert2001}%
  \BibitemOpen
  \bibfield  {author} {\bibinfo {author} {\bibfnamefont {E.}~\bibnamefont
  {Limpert}}, \bibinfo {author} {\bibfnamefont {W.~A.}\ \bibnamefont {Stahel}},
  \ and\ \bibinfo {author} {\bibfnamefont {M.}~\bibnamefont {Abbt}},\ }\href
  {\doibase 10.1641/0006-3568(2001)051[0341:lndats]2.0.co;2} {\bibfield
  {journal} {\bibinfo  {journal} {{BioScience}}\ }\textbf {\bibinfo {volume}
  {51}},\ \bibinfo {pages} {341} (\bibinfo {year} {2001})}\BibitemShut
  {NoStop}%
\bibitem [{\citenamefont {Ghadge}\ \emph {et~al.}(2010)\citenamefont {Ghadge},
  \citenamefont {Killingback}, \citenamefont {Sundaram},\ and\ \citenamefont
  {Tran}}]{Ghadge2010}%
  \BibitemOpen
  \bibfield  {author} {\bibinfo {author} {\bibfnamefont {S.}~\bibnamefont
  {Ghadge}}, \bibinfo {author} {\bibfnamefont {T.}~\bibnamefont {Killingback}},
  \bibinfo {author} {\bibfnamefont {B.}~\bibnamefont {Sundaram}}, \ and\
  \bibinfo {author} {\bibfnamefont {D.~A.}\ \bibnamefont {Tran}},\ }\href
  {\doibase 10.1080/17445760903429963} {\bibfield  {journal} {\bibinfo
  {journal} {International Journal of Parallel, Emergent and Distributed
  Systems}\ }\textbf {\bibinfo {volume} {25}},\ \bibinfo {pages} {223}
  (\bibinfo {year} {2010})}\BibitemShut {NoStop}%
\bibitem [{\citenamefont {Sinatra}\ \emph {et~al.}(2015)\citenamefont
  {Sinatra}, \citenamefont {Deville}, \citenamefont {Szell}, \citenamefont
  {Wang},\ and\ \citenamefont {Barabasi}}]{Sinatra2015}%
  \BibitemOpen
  \bibfield  {author} {\bibinfo {author} {\bibfnamefont {R.}~\bibnamefont
  {Sinatra}}, \bibinfo {author} {\bibfnamefont {P.}~\bibnamefont {Deville}},
  \bibinfo {author} {\bibfnamefont {M.}~\bibnamefont {Szell}}, \bibinfo
  {author} {\bibfnamefont {D.}~\bibnamefont {Wang}}, \ and\ \bibinfo {author}
  {\bibfnamefont {A.-L.}\ \bibnamefont {Barabasi}},\ }\href
  {https://doi.org/10.1038/nphys3494} {\bibfield  {journal} {\bibinfo
  {journal} {Nature Physics}\ }\textbf {\bibinfo {volume} {11}},\ \bibinfo
  {pages} {791} (\bibinfo {year} {2015})}\BibitemShut {NoStop}%
\bibitem [{\citenamefont {Simkin}\ and\ \citenamefont
  {Roychowdhury}(2013)}]{Simkin2013}%
  \BibitemOpen
  \bibfield  {author} {\bibinfo {author} {\bibfnamefont {M.~V.}\ \bibnamefont
  {Simkin}}\ and\ \bibinfo {author} {\bibfnamefont {V.~P.}\ \bibnamefont
  {Roychowdhury}},\ }\href {https://doi.org/10.1007/s10955-012-0677-5}
  {\bibfield  {journal} {\bibinfo  {journal} {Journal of Statistical Physics}\
  }\textbf {\bibinfo {volume} {151}},\ \bibinfo {pages} {319} (\bibinfo {year}
  {2013})}\BibitemShut {NoStop}%
\bibitem [{\citenamefont {Leydesdorff}\ \emph {et~al.}(2018)\citenamefont
  {Leydesdorff}, \citenamefont {Wagner},\ and\ \citenamefont
  {Bornmann}}]{Leydesdorff2018}%
  \BibitemOpen
  \bibfield  {author} {\bibinfo {author} {\bibfnamefont {L.}~\bibnamefont
  {Leydesdorff}}, \bibinfo {author} {\bibfnamefont {C.~S.}\ \bibnamefont
  {Wagner}}, \ and\ \bibinfo {author} {\bibfnamefont {L.}~\bibnamefont
  {Bornmann}},\ }\href {https://doi.org/10.1007/s11192-018-2734-6} {\bibfield
  {journal} {\bibinfo  {journal} {Scientometrics}\ }\textbf {\bibinfo {volume}
  {116}},\ \bibinfo {pages} {623} (\bibinfo {year} {2018})}\BibitemShut
  {NoStop}%
\end{thebibliography}%
\end{document}